\documentclass[]{aa}
\usepackage{graphicx}
\usepackage{color}
\usepackage{txfonts}

\newcommand{\beq}{\begin{equation}}
\newcommand{\eeq}{\end{equation}}
\newcommand{\beqa}{\begin{eqnarray}}
\newcommand{\eeqa}{\end{eqnarray}}

\begin{document}

\title{Radio seismology of the outer solar corona}

\author{Zaqarashvili, T.V.\inst{1,4}, Melnik, V.N.\inst{2}, Brazhenko, A.I.\inst{3}, Panchenko, M.\inst{1}, Konovalenko, A.A.\inst{2},  Franzuzenko, A.V.\inst{3},  Dorovskyy, V.V.\inst{2}, Rucker, H.O.\inst{1}
}

 \institute{Space Research Institute, Austrian Academy of Sciences, Schmiedlstrasse 6, 8042 Graz, Austria\\
             \email{teimuraz.zaqarashvili@oeaw.ac.at}
              \and
             Institute of Radio Astronomy, Ukrainian Academy of Sciences of Ukraine, Chervonopraporna st., 4, 61002, Kharkiv, Ukraine\\
                       \and
             Institute of Geophysics, Myasoedova str., 27/29, 36014, Poltava,Ukraine\\
                               \and
            Abastumani Astrophysical Observatory at Ilia State University, Cholokashvili Ave. 3/5, Tbilisi, Georgia\\
      }

\date{Received / Accepted }

\abstract{Observed oscillations of coronal loops in EUV lines have been successfully used to estimate plasma parameters in the inner corona ($< 0.2 \, R_0$, where $R_0$ is the solar radius). However, coronal seismology in EUV lines fails for higher altitudes because of rapid decrease in line intensity.}{We aim to use radio observations to estimate the plasma parameters of the outer solar corona ($> 0.2 \, R_0$).}{We use the large Ukrainian radio telescope URAN-2 to observe type IV radio burst at the frequency range of 8-32 MHz during the time interval of 09:50-12:30 UT in April 14, 2011. The burst was connected to C2.3 flare, which occurred in AR 11190 during 09:38-09:49 UT. The dynamic spectrum of radio emission shows clear quasi-periodic variations in the emission intensity at almost all frequencies.}{Wavelet analysis at four different frequencies (29 MHz, 25 MHz, 22 MHz and 14 MHz) shows the quasi-periodic variation of emission intensity with periods of $\sim$ 34 min and $\sim$ 23 min. The periodic variations can be explained by the first and second harmonics of vertical kink oscillation of transequatorial coronal loops, which were excited by the same flare. The apex of transequatorial loops may reach up to 1.2 $R_0$ altitude. We derive and solve the dispersion relation of trapped MHD oscillations in a longitudinally inhomogeneous magnetic slab. The analysis shows that a thin (with width to length ratio of 0.1), dense (with the ratio of internal and external densities of $\geq 20$) magnetic slab with weak longitudinal inhomogeneity may trap the observed oscillations. Seismologically estimated Alfv\'en speed inside the loop at the height of $\sim$ 1 $R_0$ is $\sim$ 1000 km s$^{-1}$. Then the magnetic field strength at this height is estimated as $\sim$ 0.9 G. Extrapolation of magnetic field strength to the inner corona gives $\sim$ 10 G at the height of 0.1 $R_0$.}{Radio observations can be successfully used for sounding of the outer solar corona, where EUV observations of coronal loops fail. Therefore the radio seismology of outer solar corona is complementary to the EUV seismology of the inner corona.}


\titlerunning{Radio seismology of solar corona}

\authorrunning{Zaqarashvili et al.}

\maketitle

\section{Introduction}

Magnetohydrodynamic (MHD) waves and oscillations are ubiquitous in the solar corona (Nakariakov \& Verwichte \cite{nakariakov4}). The waves may heat the ambient plasma and accelerate solar wind particles (Aschwanden \cite{Aschwanden2004}). Observations show the existence of almost all MHD modes (fast kink, fast sausage and slow magneto-acoustic waves) in solar corona (Aschwanden et al. \cite{aschwanden1}, Nakariakov et al. \cite{nakariakov1}, Ofman et al. \cite{ofm1}, Nakariakov et al. \cite{Nakariakov2003}, Wang et al. \cite{wang1,wang3}, Wang \& Solanki \cite{wang2}, Srivastava et al. \cite{Srivastava} ). Oscillations are mainly observed through imaging observations, but their spectroscopical indications are also possible. For example, purely incompressible torsional oscillation can be observed in coronal loops as periodic variation of spectral line broadening (Zaqarashvili \cite{Zaqarashvili2003}).

Observed MHD oscillations can be used to estimate plasma parameters in the solar corona using known theoretical properties of MHD waves (Roberts et al. \cite{roberts1984}). This method, so called {\it coronal seismology} (Nakariakov \& Ofman \cite{Nakariakov2001}), was effectively used to determine magnetic field strength, stratification and other plasma properties of coronal loops (Arregui et al. \cite{Arrequi2007}, Van Doorsselaere et al. \cite{Van Doorsselaere2007}, McEwan et al. \cite{McEwan2008}, Andries et al. \cite{Andries2009}, Verth et al. \cite{Verth2010}, Arregui and Asensio Ramos \cite{Arrequi2011}), spicules (Zaqarashvili et al. \cite{Zaqarashvili20071}, Zaqarashvili and Erd\'elyi \cite{Zaqarashvili2009}, Verth et al. \cite{Verth2011}) and prominences (Arregui et al. \cite{Arrequi2012} and references therein). The oscillations are mainly observed in the inner corona, where intensity of EUV lines allows to detect active region  coronal loops. In the outer corona, the oscillations are not yet detected because of EUV line intensity decrease with altitude.

\begin{figure}
\vspace*{1mm}
\begin{center}
\includegraphics[width=9cm]{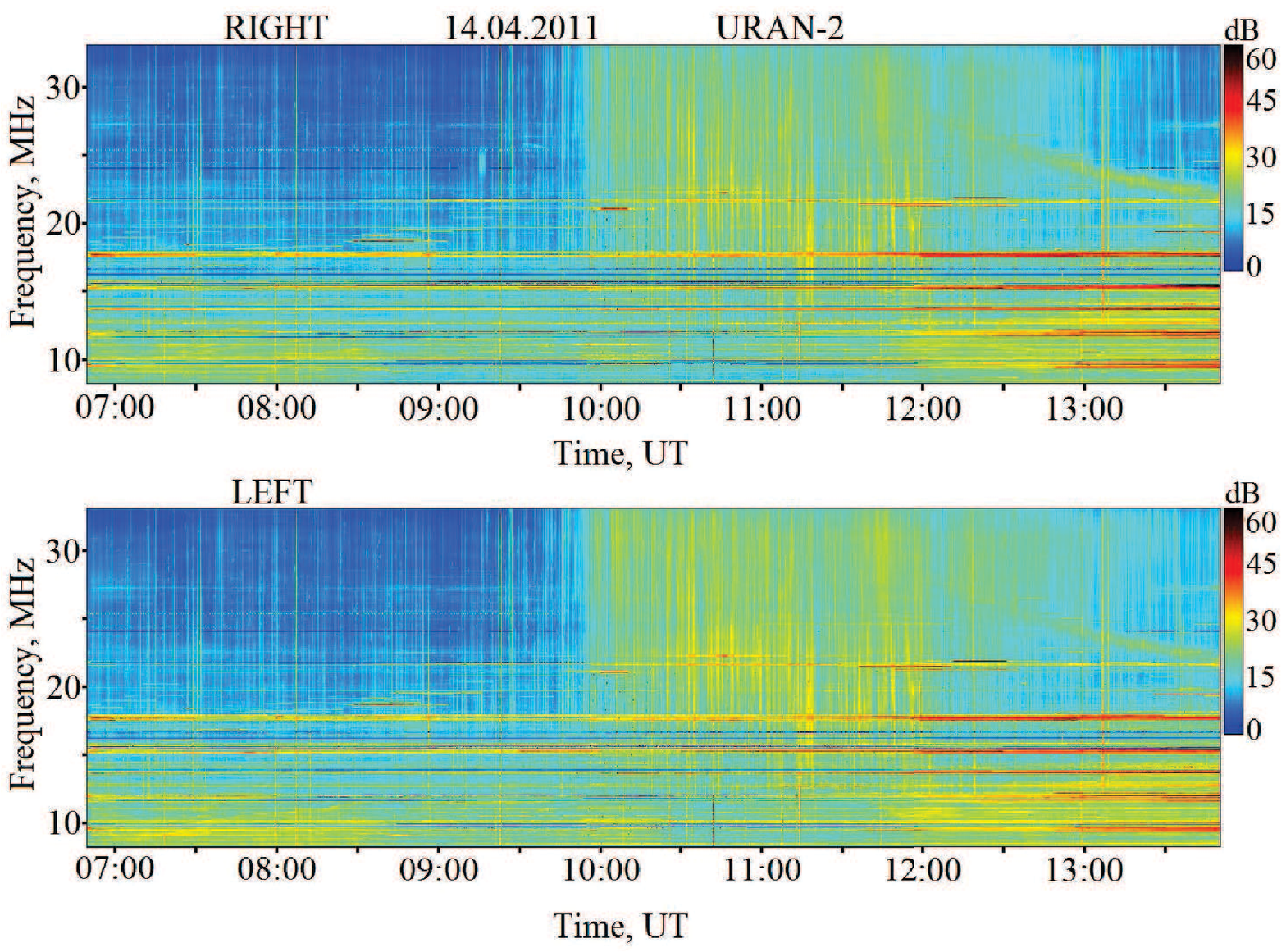}
\includegraphics[width=9.5cm]{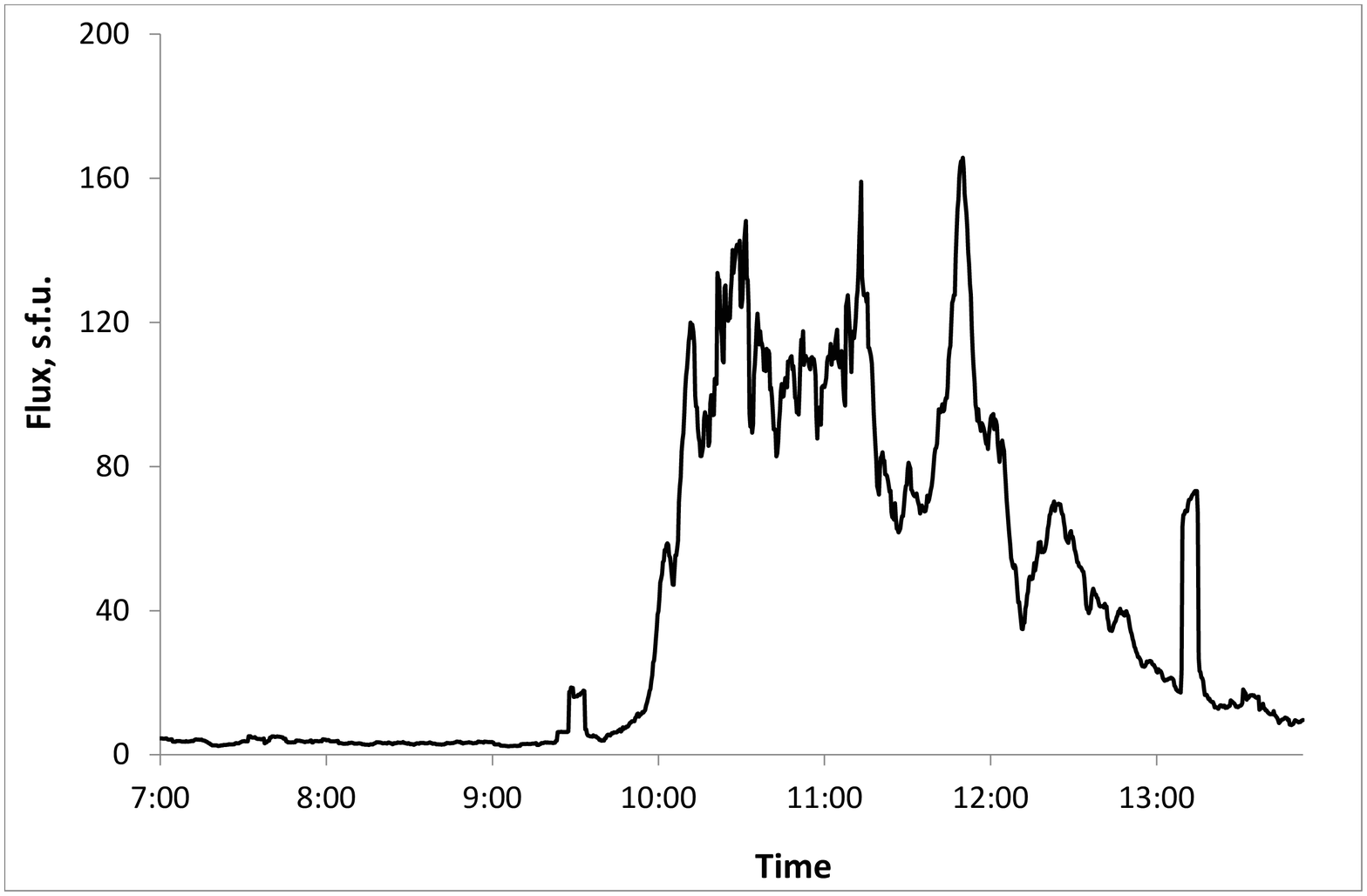}
\end{center}
\caption{Dynamic spectrum of radio emission (right and left polarization) observed by URAN-2 on April 14, 2011 (upper panel) and time profile of this radio emission at 29.7 MHz (lower panel).}
\end{figure}

Besides the EUV part of spectrum, the solar corona can be observed in the radio frequency band, which is very important to study the rapid oscillations of the solar atmosphere owing to high temporal resolution. High-frequency oscillations ($<$ 1 min) are frequently detected in the radio observations (Aschwanden et al. \cite{Aschwanden20041}, see Table 1 in the paper). According to Aschwanden et al. (\cite{Aschwanden20041}) most of the reported oscillations fall in the frequency range of 100-1000 MHz corresponding to the electron densities of 10$^8$-10$^{10}$ cm$^{-3}$. Therefore, the radio emission is originated from the inner corona and consequently the oscillations can be interpreted as fast sausage modes in short low-laying coronal loops (Nakariakov et al. \cite{Nakariakov2003}, Aschwanden et al. \cite{Aschwanden20041}). Long period modulation (with 5-10 min) in the microwave range of $>$1 GHz  can be explained by the kink oscillations of coronal loops (Khodachenko et al. \cite{Khodachenko2011}). Recent high-resolution data obtained with the Nobeyama Radioheliograph show the frequency drift of 3-min oscillations, which can be explained by the dispersive evolution of upward propagating wave pulses (Sych et al. \cite{Sych2012}). Hence, the radio observations are very important to study the plasma properties in the inner corona.

On the other hand, radio observations in the frequency range of 10-100 MHz are very important in the outer corona.
Flare related activity at decameter wavelengths is manifested in great variety of bursts: type II-IV bursts, drifting pairs, S-bursts,
spikes bursts in absorption etc (Melnik et al. \cite{Melnik2010}). Some of them are continuations of the same bursts from decimeter and meter bands such
as type III bursts, type II bursts, type IV bursts, spikes. There are some bursts, which are observed in the long meter and decameter range only, namely
type IIIb bursts, drifting pairs, S-bursts. Plasma mechanism of radio emission is usually used for explanation of these bursts
(Ginzburg and Zhelezniakov \cite{Ginzburg1958}). This mechanism has two stages. The first stage is the generation of Langmuir waves by fast
electrons and the second stage is the emission of electromagnetic waves by the Langmuir waves. In this mechanism, the frequency of radio emission
equals the local plasma frequency or double local frequency. Then the radio emission at given frequency is originated in the plasma at corresponding height above the solar surface. In this paper we use the decameter band (8-32 MHz), therefore the radio emission at
these frequencies is generated at heights $\geq$ 1 $R_0$, where $R_0$ is the solar radius, depending on the density structure of the solar corona. Type IV bursts at decameter wavelength were first observed by the radio telescope UTR-2 during observational campaign in 2002-2004 (Melnik et al. \cite{Melnik2008}). The duration of the bursts was from 1.5 to several hours and fluxes were 10-10$^3$ s.f.u. Decameter type IV bursts were usually accompanied by coronal mass ejections (CME's) but sometimes there was no visible CME during type IV burst. Specific features of decameter type IV bursts were fine structure in the form of fiber-bursts and oscillations of radio emission intensity. Largest periods of these oscillations were tens of minutes. It was suggested that the radio emission of type IV bursts originates either in a coronal loop disturbed by CME or in the body of CME itself. Hence, the observed oscillations can be connected either with oscillations of CME or coronal loops. Oscillation in the intensity of type IV bursts without accompanied CME is probably connected to the coronal loop oscillations triggered by a solar flare. Then, the analysis of these oscillations allows to estimate the plasma properties in the coronal loops. Among decameter bursts there are so called inverted U- and J- bursts generated by fast electron beams, which propagate in high coronal loops (Dorovsky et al. \cite{Dorovsky2010}).
Active region coronal loops may expand up to helmet streamers owing to solar wind. Additionally, very long transequatorial loops, which connect the active regions in opposite hemispheres, may reach up to the height of one solar radius. Chase et al. (\cite{Chase1976}) presented the first observational indication of transequatorial loops. Later, Pevtsov (\cite{Pevtsov2000}) presented 87 transequatorial loops through the analysis of Yohkoh data set between 1991 October and 1998 December. The radio emission at 10-30 MHz are originated from the heights, where transequatorial and/or active region loops are located. Then, the long-period modulation of emission intensity at these frequencies can be caused by the oscillations of the loops.

Here we use the observation of type IV radio burst from the large Ukrainian radio telescope URAN-2 for the seismological estimation of magnetic field strength and other plasma parameters of coronal loops at the heights of $\geq$ 0.2 $R_0$.

\section{Observations}

Ukrainian radio telescope URAN-2 (Poltava, Ukraine) operates in the frequency range of 8-32 MHz. URAN-2 has effective square of 28000 m$^2$ and its beam angular resolution is 3.5$^0$-7$^0$(Megn et al. \cite{Megn2003}, Brazhenko et al. \cite{Brazhenko2005}). The radio telescope URAN-2 can measure polarization of radio emission.
The digital spectrometers DSPz (Ryabov et al. \cite{Ryabov2010}) registered the solar data in above mentioned whole frequency range. Type IV radio burst was detected in the frequency range of
$\sim$ 12-32 MHz during the time interval of 09:50-12:30 UT in April 14, 2011 with the frequency resolution of 4 kHz and the time resolution of 100 ms. Figure 1 shows the dynamic spectrum of solar radio emission observed
by radio telescope URAN-2 in April 14, 2011 and the time profile of the radio emission at the frequency of 29.7 MHz. Decameter
type IV burst started at 10:00, continued about 4 hours and finished approximately at 14:00. Polarization of type IV burst was high about 40-50 $\%$.
The type IV radio burst was probably connected to small C2.3 flare, which occurred in the active region AR 11190. The flare started at 09:38 UT, reached the maximum phase at 09:44 UT and ended at 09:49 UT.

\begin{figure}[t]
\vspace*{1mm}
\begin{center}
\includegraphics[width=9.5cm]{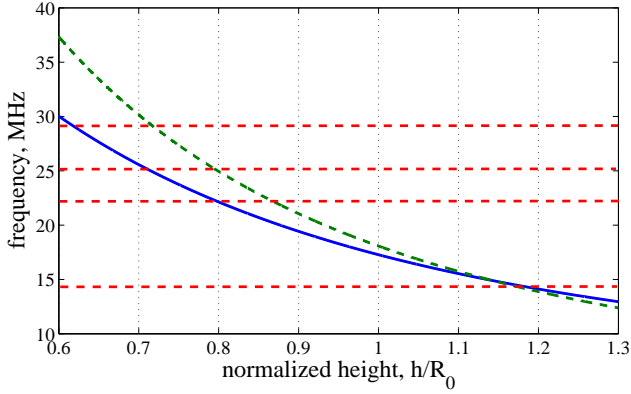}
\end{center}
\caption{Frequency of radio emission vs height above the solar surface. Blue solid line shows the dependence according to Baumbach-Allen formula of quiet Sun density structure. Green dashed line shows the dependence according to gravitationally stratified atmosphere with 1 MK temperature. Red dashed lines correspond to 29 MHz, 25 MHz, 22 MHz and 14 MHz frequencies respectively. }
\end{figure}

Type IV Radio bursts are generally produced by an emission mechanism near the plasma frequency, which depends on the electron density $n_e$ as
\begin{equation}\label{plasma-frequency}
\nu_p= 8980 \sqrt{n_e},
\end{equation}
where $\nu_p$ is in Hz and $n_e$ is in cm$^{-3}$.  Therefore, the radio emission at particular frequency is excited at particular local density of emitted plasma. The density of the solar corona generally decreases with distance, hence the radio emission at different frequencies corresponds to different heights above the solar surface.
The density structure of the quiet Sun corona can be approximated by the Baumbach-Allen formula (Aschwanden \cite{Aschwanden2004})
\begin{equation}\label{baumbach-allen}
n_e(\zeta)=10^8\left ({2.99\over {(1+\zeta)^{16}}}+{1.55\over {(1+\zeta)^{6}}}+{0.036\over {(1+\zeta)^{1.5}}} \right ) \ {\rm cm}^{-3},
\end{equation}
where $\zeta=h/R_0$ is the normalized height above the solar surface. However, the density stratification can be different in coronal loops, which
are denser than the ambient plasma. Therefore, the Baumbach-Allen formula can not be directly applied to the coronal loops.

\begin{figure*}
\vspace*{1mm}
\begin{center}
\includegraphics[width=18cm]{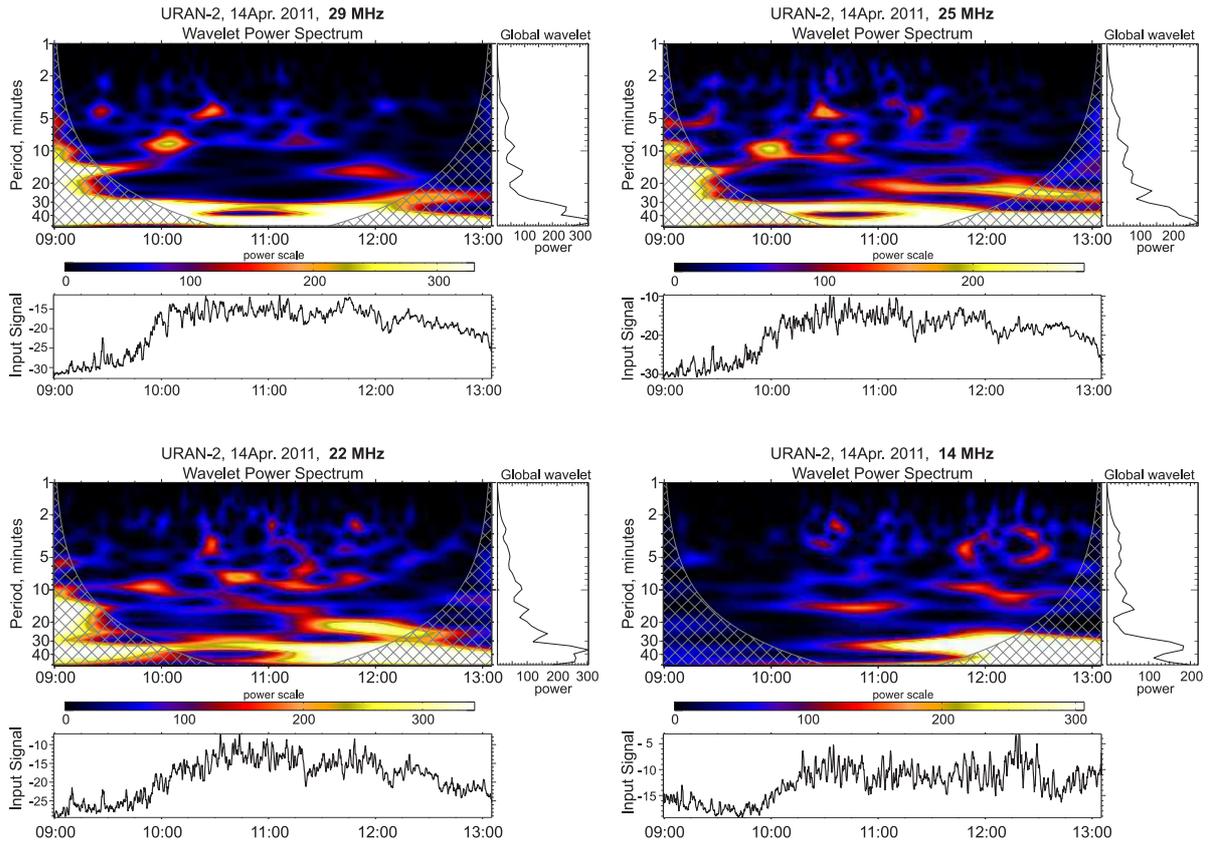}
\end{center}
\caption{Morlet wavelet power spectrum of emission intensity at 29 MHz (upper left panel), 25 MHz (upper right panel), 22 MHz (lower left panel) and 14 MHz (lower right panel) frequencies during 09:00-13:00 UT. Dashed areas outside the cone of influence (COI) indicate the regions where wavelet transform is not reliable. }
\end{figure*}

Figure 2 shows the radio emission frequency vs height above the solar surface according to Baumbach-Allen formula (Eq.~\ref{baumbach-allen}) and from the gravitationally stratified coronal loop, which is expressed by formula (see Appendix A)
\begin{equation}\label{strat}
n_e={n_{e0}}\exp\left ({-{R_0\over H_n}{\zeta\over {1+\zeta}}}\right ) \ {\rm cm}^{-3},
\end{equation}
where $H_{n}=2kT/mg$ is the density scale height near the solar surface (here $k=1.38\cdot 10^{-16}$ erg K$^{-1}$ is the Boltzmann constant, $m=1.67 \cdot 10^{-24}$ g is the proton mass, $g=2.74\cdot 10^{4}$ cm s$^{-2}$ is the gravitational acceleration near the solar surface and $T=$ 1
MK is the plasma temperature) and $n_{e0}$ is chosen in such way that it is 3 times higher than corresponding Baumbach-Allen value at the height of 0.1 $R_0$.
The curves corresponding to Baumbach-Allen formula and gravitationally stratified corona are significantly different at lower heights, but become comparable near 1 $R_0$. However, the two approaches may give quite different results. From this figure it is seen that the radio emission at 13-32 MHz frequency corresponds to the heights of 0.6-1.2 $R_0$ in the case of Baumbach-Allen formula and to the heights of 0.7-1.2 $R_0$ in the case of gravitationally stratified corona.

\section{Quasi-periodic variations in radio emission intensity}

Dynamic spectrum of radio emission (Fig. 1) shows that the variation of emission intensity has quasi-periodic behavior in time at all frequencies, which indicates that the physical parameters of the solar corona (density of energetic electron beams, local plasma density, magnetic field strength and/or direction etc.) undergo quasi-periodic variations. In order to understand the reason of the variations, one needs to estimate statistically significant periods in the time series.

\subsection{Wavelet analysis}

We use wavelet analysis to study the temporal variation of emission intensity at different frequencies. Fig. 3 shows the wavelet spectrum of
emission intensity during the interval 09:00-13:00 UT at four different frequencies, namely at 29 MHz, 25 MHz, 22 MHz and 14 MHz.

Power spectrum of 29 MHz emission shows a clear peak at 33.5 min persisted during almost whole interval. Power spectra of 25 and 22 MHz emissions show peaks at 34 $\pm$ 2 min and 31.5 MHz respectively. The oscillations also persist during long time, however sometimes they are outside of COI. There is an additional indication of $\sim$ 23 min periodicity in both plots. It should be mentioned that the power spectra of 29 MHz, 25 MHz and 22 MHz emissions, which correspond to 0.6-0.8 $R_0$ heights for Baumbach-Allen formula and to 0.7-0.9 $R_0$ for the gravitationally stratified corona, are very similar. On the other hand, the power spectrum of 14 MHz is different, but the periodicity near 34 min is still present. Therefore, the oscillations with $\sim$ 34 min are present at all frequencies with a significant power. There is an indication of longer period oscillation with $\sim$ 45 min at 29 MHz and 25 MHz frequencies, but it is at the limit of convincing significance of the wavelet power spectrum. There are shorter period oscillations which exist only within short intervals, but they have much lower significance, therefore we do not consider them here.

\begin{table*}[ht!]
\caption{Oscillation periods at different frequencies of radio emission obtained by wavelet analysis}
\label{tab:1}       
\begin{tabular}{llll}
\hline\noalign{\smallskip}
29 MHz & 25 MHz  & 22 MHz & 14 MHz \\
\noalign{\smallskip}\hline\noalign{\smallskip}
47 min (?) & 40 min (?) &  &    \\
33.5 min & 34 min & 31.5 min & 34.6 \\
 & 23.4 min & 23 min &  \\

\noalign{\smallskip}\hline
\end{tabular}
\end{table*}


Table 1 summarizes the results of wavelet analysis. The period of $\sim$ 34 min has stronger probability and it is seen at all frequencies during long time interval. The period of $\sim$ 23 min period is also clearly seen at the frequencies of 25 MHz and 22 MHz.

\subsection{Possible mechanism: coronal loop oscillations}

Quasi-periodic variations in the radio emission intensity can be caused by several mechanisms. Energetic electron beams generated during the solar flare may excite Langmuir oscillations in coronal loops, which may emit the radio emission at observed frequencies as type IV burst. Variation of electron beam density may influence the amplitude of Langmuir oscillations and consequently the intensity of radio emission. This requires the variation of cross section of coronal loop where the electron beam propagates. Therefore, the oscillation of magnetic tube cross section may lead to the quasi-periodic variation of beam density and consequently to the variation of emission intensity. Periodic variation of the angle between the magnetic field direction and the line of sight may also lead to the observed oscillation. Therefore, the transverse oscillation of coronal loop can be also considered as a possible mechanism for the observations.

It is seen from Figure 2 that the emissions at 29 MHz, 25 MHz, 22 MHz and 14 MHz frequencies correspond to $\sim 0.6\, R_0$, $\sim 0.7\, R_0$, $\sim 0.8\, R_0$ and $\sim 1.2\, R_0$ heights for Baumbach-Allen formula and $\sim 0.7\, R_0$, $\sim 0.8\, R_0$, $0.9\, R_0$ and $1.2\, R_0$ heights for gravitationally stratified corona. Active region coronal loops usually reach 0.1 $R_0$ height in coronal EUV lines, but in reality they may expand up to the higher heights. On the other hand, very long transequatorial loops, with the length of $10^0$-$75^0$ heliographic degrees ($1^0$ corresponds to $\approx 1.2\cdot10^4$ km), reach higher heights (Pevtsov \cite{Pevtsov2000}). The apex of transequatorial loops can be located at observed heights, therefore MHD oscillations of the loops may produce the observed variations.

Small C2.3 flare, which probably triggered the type IV radio burst, occurred in AR 11190. Figure 4 shows the magnetogram of solar disc at the photospheric level during the observed event obtained by Helioseismic and Magnetic Imager (HMI) (Schou et al \cite{Schou2012}) on the board of Solar Dynamic Observatory (SDO) (Pesnell et al. \cite{Pesnell2011}). The active region AR 11190 is located at the central part of the northern hemisphere, while the active region AR 11187 is located just opposite in the southern hemisphere. Therefore, the two active regions could be connected by transequatorial loop system. Transequatorial loops may connect two other active regions of the northern and southern hemispheres, namely AR 11186 and AR 11188. The distance between AR11190 and AR11187 is about $40^0$, which is around $\approx 4.8\cdot10^5$ km i.e. 0.7 $R_0$. On the other hand, the distance between AR11186 and AR11188 is about $50^0$, which is around $\approx 6\cdot10^5$ km i.e. 0.86 $R_0$. In the case of semi-circular shape, the transequatorial loops may reach up to 0.35 $R_0$ in the first case and 0.43 $R_0$ in the second one. In the case of sheared magnetic arcade, the transequatorial loops may expand much higher up to the observed heights.

Figure 5 shows the solar corona in Fe IX 171 \AA\ line during the observed events obtained by Atmospheric Imaging Assembly (AIA) on SDO (Lemen et al. \cite{Lemen2012}). Transequatorial loop system, which connects the active regions AR11190 and AR11187 at lower heights, is seen. The loop system is probably expanded to much higher heights, but it is not seen in EUV lines.

Solar flare in AR 11190 may excite energetic electron beams along transequatorial loops, which may trigger the Langmuir oscillations producing observed radio emission. The solar flare may also trigger the MHD oscillations of transequatorial loops, which modulate the electron beams density and lead to the observed quasi-periodicity in radio intensity (Fig. 6).

\section{Seismology of the outer corona}

Observed oscillation periods can be used to estimate the plasma parameters of the outer solar corona using theoretical oscillation spectrum of MHD waves in coronal loops. Analytical dispersion relations for MHD waves for straight homogeneous magnetic tubes was developed at the beginning of 80th by Edwin and Roberts (\cite{Edwin1983}). Recent progress in coronal observations owing to Transition Region And Coronal Explorer (TRACE) triggered numerous theoretical studies for coronal loops with longitudinal density inhomogeneity (D\'iaz et al. \cite{diaz1}, Van Doorsselaere et al. \cite{van1}, Andries et al. \cite{andries1}, Dymova \& Ruderman \cite{dymova}, Donnelly et al.
\cite{donnelly}, Zaqarashvili and Murawski \cite{Zaqarashvili20072}), for the loops with longitudinally inhomogeneous magnetic field (Verth and Erd\'elyi \cite{verth}, Pascoe et al. \cite{Pascoe2009}) and for curved loops (Brady \& Arber \cite{brady}, Selwa et al. \cite{selwa, Selwa2007}, Verwichte et al. \cite{verwichte1}, Di\'az et al. \cite{diaz2006}, Terradas et al. \cite{terradas}, Gruszecki et al. \cite{Gruszecki}).

Slow magneto-acoustic oscillations can be ruled out as the reason of observed modulation of radio emission. The fundamental standing slow wave in the coronal loop with the apex at 0.8 $R_0$ should have a period of 3-6 hr (for the sound speed of 150-300 km s$^{-1}$), which is too long compared to the observed periodicity. Kink, sausage and torsional oscillations could be responsible for the observed variation of radio emission. Kink and torsional oscillations change the direction of loop magnetic field periodically, therefore they may lead to the variation of radio emission (Khodachenko et al. \cite{Khodachenko2011}). On the other hand, sausage oscillations lead to the variation of the loop cross-section and may modulate the density of electron beams, which in turn may lead to the variation of radio emission intensity. Kink oscillation in vertical plane, so called vertical kink oscillations, may also lead to the significant variation of loop cross-section and thus can lead to the similar effect of sausage oscillations (Aschwanden and Schrijver \cite{aschwanden2011}).

The highest altitude of emitted radio emission (corresponding to the 14 MHz frequency) is located at 1.2 $R_0$ (see Fig. 2). It may correspond to the apex of transequatorial coronal arcade. The mean length of the arcade can be estimated as
\beqa
2L=\pi \cdot 1.2 \cdot R_0 \approx 2.64 \cdot 10^6 \, {\rm km}
\label{eq:length}
\eeqa
for both Baumbach-Allen formula and
for the gravitationally stratified corona.

In the next subsections we will use MHD oscillations of coronal loops for the estimation of plasma parameters in the outer corona using observed oscillation periods.

\subsection{Kink oscillations}

Kink waves are transverse oscillations of magnetic tubes and the phase speed for a straight homogeneous tube can be written as
\beqa
c_k=\sqrt{{{\rho_{\rm 0} V^2_{\rm A0}+\rho_{\rm e} V^2_{\rm Ae}}\over {\rho_{\rm 0}+\rho_{\rm e}}}}=V_{\rm A0}\sqrt{{{2}\over {1+\rho_{\rm e}/\rho_{\rm 0}}}},
\label{eq:ck}
\eeqa
where ${\rho}_{\rm 0}$ (${\rho}_{\rm e}$) is the plasma density inside (outside) the loop and $V_{\rm A0}=B_{\rm 0}/\sqrt{4\pi{\rho}_{\rm 0}}$ ($V_{\rm Ae}=B_{\rm 0}/\sqrt{4\pi{\rho}_{\rm e}}$) is the Alfv{\'e}n speed inside (outside) the loop. Note that the magnetic field strength is the same inside and outside the loop, which is a good approach in the solar corona. The horizontal transverse oscillation of coronal loops, i.e. parallel to the surface, changes the direction of loop magnetic field, but the density variation remains small. On the other hand, vertical kink oscillation may lead to the density variation due to the variation of loop length (Aschwanden and Schrijver \cite{aschwanden2011}).

\begin{figure}
\vspace*{1mm}
\begin{center}
\includegraphics[angle=90, width=11cm]{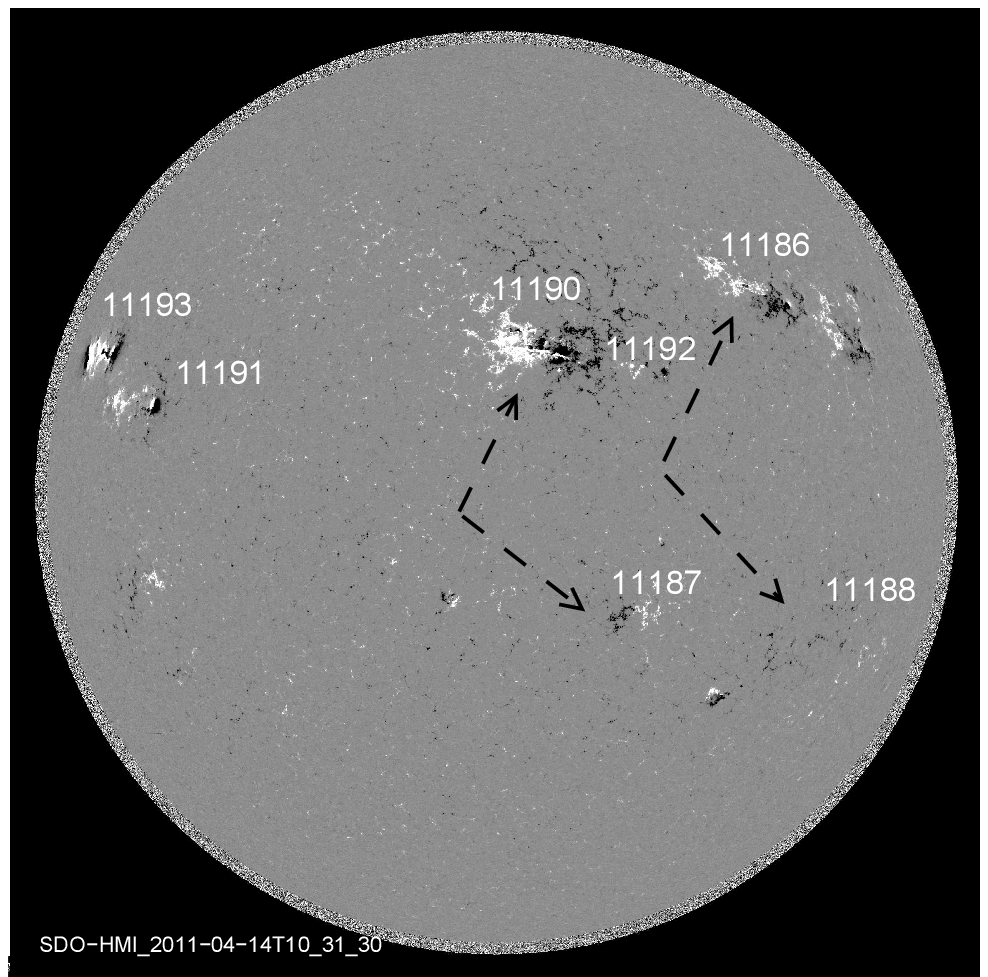}
\end{center}
\caption{Solar photospheric magnetogram at 10:31:30 UT in April 14, 2011 from SDO/HMI. Active regions are symmetrically located in northern and southern hemispheres and can be connected by transequatorial loops. The dashed arrows indicate the possibilities of transequatorial loop systems.}
\end{figure}

Suppose that the observed oscillation is caused by the first harmonic of kink waves. Then the Alfv\'en speed inside the loop can be estimated as
\beqa
V_{\rm A0}={{4L}\over {T_{\rm obs}}}\sqrt{{{1+\rho_{\rm e}/\rho_{\rm 0}}\over {2}}}\approx 2000 \, {\rm km}{\cdot}{\rm s}^{-1},
\label{eq:alfk}
\eeqa
where the observed period of $T_{\rm obs}=34$ min, the loop length of $2L=2.64 \cdot 10^6$ km and density ratio of $\rho_{\rm e}/\rho_{\rm 0}=1/5$ are used.


\subsection{Sausage oscillations}

Sausage waves modify the loop cross-section, therefore they may lead to the variation of beam density and consequently the radio emission intensity.
However, the long wave length sausage waves are evanescent in magnetic tubes (Edwin \& Roberts \cite{Edwin1983}). The expression of cut-off wave number can be written in the cold plasma approximation as (Roberts et al. \cite{roberts1984}, Nakariakov et al. \cite{Nakariakov2003}, Aschwanden et al. \cite{Aschwanden20041})
\beqa
k_c={{j_{0}}\over {x_0}}\sqrt{{{1}\over {\rho_{\rm 0}/\rho_{\rm e}-1}}},
\label{eq:cut}
\eeqa
where $x_0$ is the tube radius and $j_0=2.4$ is the first zero of Bessel function. For the density ratio of $\rho_{\rm 0}/\rho_{\rm e}=5$, this expression leads to
\beqa
k_c x_0\approx 1.2.
\label{eq:cut1}
\eeqa
The normalized wave numbers for sausage waves are
\beqa
k_s x_0= {{2\pi x_0}\over {4L}}(n+1), \, n=0,1,2 \ldots,
\label{eq:cut2}
\eeqa
where $n=0$ corresponds to the first harmonic, $n=1$ corresponds to the second etc.

Figure 7 shows the dependence of normalized wave number $k_s x_0$ of first three harmonics on width to length ratio (aspect ratio) of magnetic tube. It is seen that the first harmonic is trapped only for thick loops: the loop with the density contrast $d=\rho_{\rm 0}/\rho_{\rm e}$ of 5, 10 and 20 require the aspect ratio of $\sim$ 0.75, $\sim$ 0.5 and $\sim$ 0.35, respectively. For the second harmonic the same density contrasts require $\sim$ 0.38, $\sim$ 0.25 and $\sim$ 0.18. The third harmonic yields $\sim$ 0.25, $\sim$ 0.18 and $\sim$ 0.12. Therefore, only the third harmonic can be trapped in reasonable values of density contrast and aspect ratio.

The upper limit of phase speed for sausage waves is the external Alfv\'en speed, $V_{\rm Ae}$. If one supposes that the third harmonic is responsible for the observed variation of radio emission, then the external Alfv\'en speed can be estimated as
\beqa
V_{\rm Ae}={{4L}\over {3T_{\rm obs}}}\approx 840 \, {\rm km}{\cdot}{\rm s}^{-1},
\label{eq:alfk}
\eeqa
where the observed period of $T_{\rm obs}=34$ min and the loop length of $2L=2.64 \cdot 10^6$ km  are used. The internal Alfv\'en speed can be estimated from the suggestive density ratio of $d=\rho_{\rm 0}/\rho_{\rm e}=5$ as
\beqa
V_{\rm A0}\approx 375 \, {\rm km}{\cdot}{\rm s}^{-1}.
\label{eq:alfk}
\eeqa
It is a rather small value of internal Alfv\'en speed at these heights.

On the other hand, if the third harmonic of sausage waves is represented by the period of $\sim$ 23 min, which has significant power in the wavelet power spectrum, then the Alfv\'en speed outside the loop can be estimated as
\beqa
V_{\rm Ae}\approx 1275 \, {\rm km}{\cdot}{\rm s}^{-1},
\label{eq:alfk}
\eeqa
which gives for the internal Alfv\'en speed as
\beqa
V_{\rm A0}\approx 570 \, {\rm km}{\cdot}{\rm s}^{-1}.
\label{eq:alfk}
\eeqa
which seems to be a small value again.

The estimation of Alfv\'en speed is based on the model of homogeneous coronal loops. However, the transequatorial loops are located in higher corona, therefore the Alfv\'en speed may vary along the loop axis. Therefore, the oscillations in the loops with longitudinally inhomogeneous Alfv\'en speed should be studied. Oscillations in coronal loops with step-like density profile are well studied (Di\'az et al. \cite{diaz1}; Dymova \& Ruderman \cite{dymova}; Donnelly et al. \cite{donnelly}). However, it is desirable to consider a smooth profile of Alfv\'en speed, therefore we use a smooth profile adopted by Zaqarashvili and Murawski (\cite{Zaqarashvili20072}).

\subsection{Oscillations of coronal loops with longitudinally inhomogeneous Alfv\'en speed}

\begin{figure}
\vspace*{1mm}
\begin{center}
\includegraphics[width=11cm]{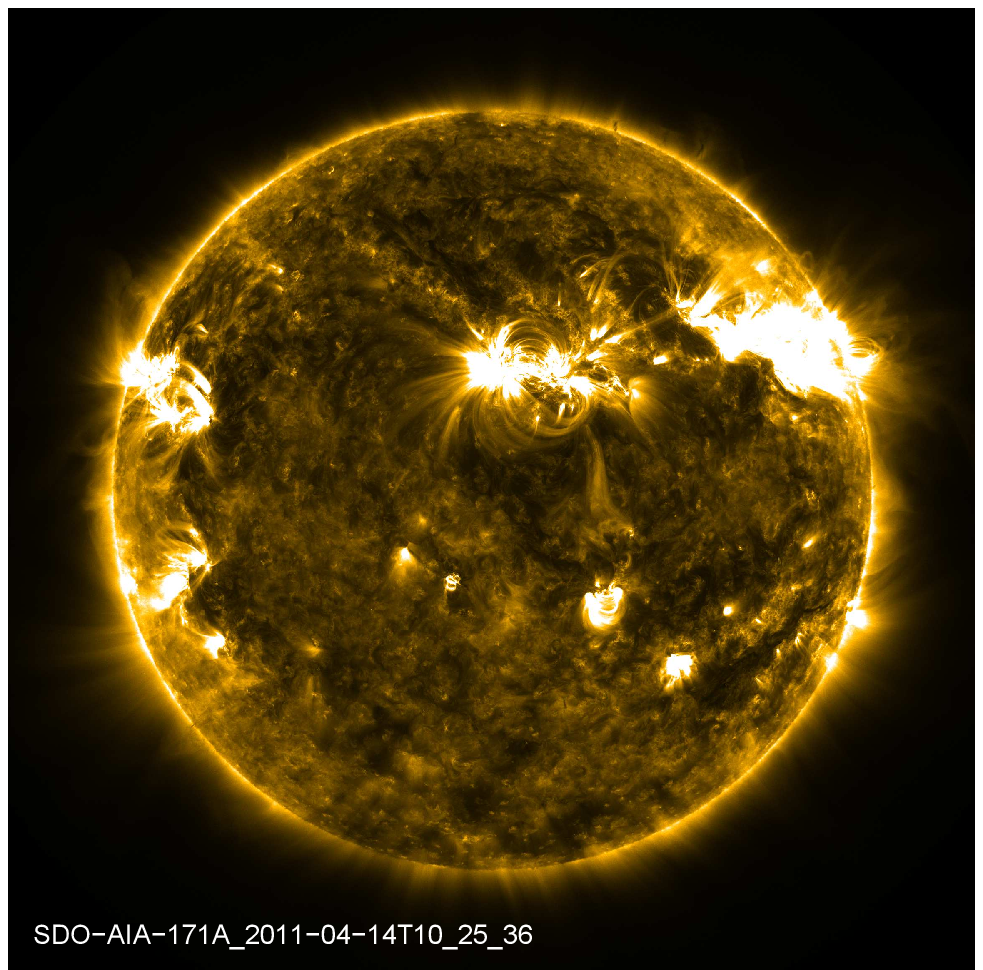}
\end{center}
\caption{Solar corona as seen in 171 \AA\ line at 10:25:31 UT in April 14, 2011 obtained by AIA/SDO.}
\end{figure}

Since slow magneto-acoustic waves are probably not relevant to our observations, we implement a zero plasma-$\beta$ approximation and take into account small amplitude perturbations. We use Cartesian coordinate system ($x,z$), therefore the considered magnetic structure is rather a slab than a loop. Consideration of cylindrical magnetic tube makes the oscillation spectrum more comprehensive, but the slab model is also a good approximation for our purposes. The equation governing the linear perturbations in the system can be
written after Fourier analysis with $\exp(-i\omega t)$ as
\beq
{{\partial^2 u_x}\over {\partial x^2}}  + {{\partial^2 u_x}\over {\partial
z^2}} + {\omega^2 \over V^2_{\rm A}}u_x=0, \,
\label{eq:main}
\eeq
where $u_x$ is the velocity perturbation, $V_A=B_{\rm 0}/\sqrt{4\pi{\varrho}}$ is the
Alfv{\'e}n speed, ${\varrho}$ is the plasma density, $B_0$ is the magnetic field strength and $\omega$ is a wave frequency. Eq. ~(\ref{eq:main}) accompanied by appropriate boundary conditions, consists a classical boundary-value problem.


We consider a straight dense coronal loop that is embedded in a rarified plasma. The low plasma-$\beta$ condition in the solar corona implies that
the enhanced hydrodynamic pressure due to the dense and hot plasma inside a coronal loop can be easily compensated by very small difference in the magnetic field strength inside and outside the loop. Alfv{\'e}n speed inside and outside the loop is given as

\beq
{V}_{\rm A0}\left (1+ {\alpha^2}{z^2\over L^2}\right )^{-1/2},\,\, {\rm for}\,\, x<x_0,
\label{eq:rhoi}
\eeq
\beq
{V}_{\rm Ae}\left (1+
{{\varrho}_{\rm 00}\over \varrho_{\rm 0e}}{\alpha^2}{z^2\over
L^2}\right )^{-1/2}\, \, {\rm for}\,\, x>x_0,
\label{eq:rhoe}
\eeq
where $V_{\rm A0}=B_{\rm 0}/\sqrt{4\pi{\varrho}_{\rm 00}}$ ($V_{\rm Ae}=B_{\rm 0}/\sqrt{4\pi{\varrho}_{\rm 0e}}$) is the internal (external) Alfv{\'e}n speed at the loop apex ($z=0$), ${\varrho}_{00}$ and ${\varrho}_{0e}$ are the corresponding densities and
$\alpha$ defines the strength of the inhomogeneity. The profile states that the Alfv{\'e}n speed is smaller at loop footpoints and becomes higher at the loop apex.

In order to find a dispersion relation of possible oscillation modes in the system one can find the solutions of Eq. ~(\ref{eq:main}) inside and outside the tube, which satisfy the closed boundary conditions at the loop footpoints, and then merge the solutions at the tube boundary (Roberts \cite{Roberts1981}).

\begin{figure}[t]
\vspace*{1mm}
\begin{center}
\includegraphics[width=7cm]{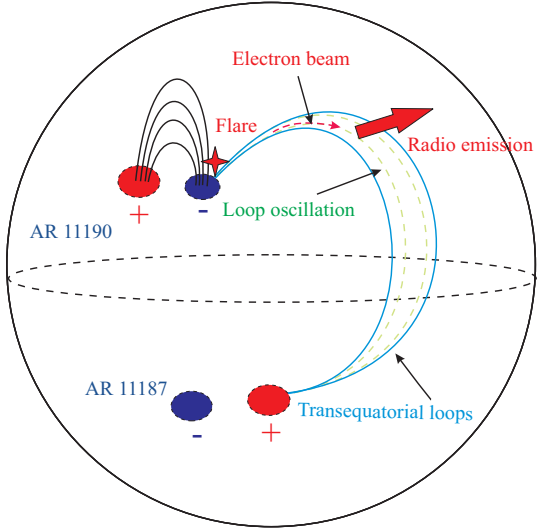}
\end{center}
\caption{Schematic picture of radio emission from a transequatorial coronal loop after a solar flare.}
\end{figure}

The velocity and the total pressure perturbations should be continuous at the
loop boundary $x=x_{\rm 0}$,
%
\beq [u_x]=0, \,\, \left [P={{B_0 b_z}\over {4\pi}}\right ]=0\, . \label{eq:at_boundary} \eeq
Here $b_z$ is the $z$ component of magnetic field perturbation,  and $[f]$ indicates the jump of
$f$ across the boundary. We implement line-tying conditions at loop
foot-points
%
 \beq
u_x(z=\pm L)=0\, .
\label{eq:foot_points}
\eeq

Inside and outside the loop, Eq.~(\ref{eq:main}) has the following form:
\beq
{{\partial^2 u_{xi}}\over {\partial x^2}}  + {{\partial^2 u_{xi}}\over {\partial
z^2}} + \left [ {\omega^2 \over V^2_{\rm
A0}} +\alpha^2{{\omega^2 } \over V^2_{\rm A0}}{z^2\over L^2} \right ] u_{xi} =0, \,
\label{eq:inside}
\eeq
\beq
{{\partial^2 u_{xe}}\over {\partial x^2}}  + {{\partial^2 u_{xe}}\over {\partial
z^2}} + \left [ {\omega^2 \over V^2_{\rm
Ae}} +\alpha^2{{\omega^2 } \over V^2_{\rm A0}}{z^2\over L^2} \right ] u_{xe} =0, \,
\label{eq:outside}
\eeq
where $u_{xi}$ ($u_{xe}$) is the transverse velocity perturbation inside (outside) the loop.

We use the method of separation of variables in order to solve Eqs.~(\ref{eq:inside})-(\ref{eq:outside}). The two equations are similar, so it is enough to solve Eq.~(\ref{eq:inside}). Eq.~(\ref{eq:outside}) can then be solved analogously.
Setting $u_{xi}(x,z)=\Phi_i(x)\Psi_i(z)$ we get from Eq.~(\ref{eq:inside})
\beq {{d^2 \Psi_i}\over
{d z^2}} + \alpha^2{{\omega^2 } \over V^2_{\rm A0}}{z^2\over
L^2}\Psi_i= -k^2 \Psi_i,
\label{eq:Psi}
\eeq
\beq
{{d^2 \Phi_i}\over
{d x^2}}
+ \left [ {\omega^2 \over V^2_{\rm A0}} - k^2\right ]\Phi_i=0\, ,
\label{eq:Phi}
\eeq
where $k^2$ is the separation constant.
Boundary conditions (\ref{eq:foot_points}) are now rewritten as
\beq
\Psi_i(z=\pm L)=0.
\label{eq:Psibound}
\eeq
Equations~(\ref{eq:Psi}) and (\ref{eq:Psibound}) consist the well-known Sturm-Liouville problem,
which requires to find the solutions of Eq.~(\ref{eq:Psi}), that satisfy the
boundary condition (\ref{eq:Psibound}).

\begin{figure}[t]
\vspace*{1mm}
\begin{center}
\includegraphics[width=9cm]{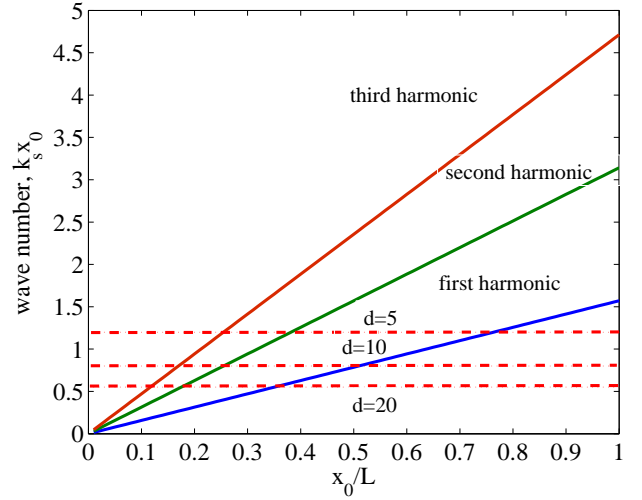}
\end{center}
\caption{Normalized wavenumber of first three harmonics of sausage waves (blue - first harmonic, green - second harmonic and dark red - third harmonic) vs width to length ratio of magnetic tube from Eq.~(\ref{eq:cut2}). Red dashed lines show the cut-off wavenumbers for three different values of the ratio of internal and external densities, $d=\rho_{\rm 0}/\rho_{\rm e}$=5, 10 and 20.}
\end{figure}

With the use of the notations (Zaqarashvili and Murawski \cite{Zaqarashvili20072})
\beq \xi \equiv \sqrt{{2\alpha \omega}\over {V_{\rm A0}L}}z,
\hspace{3mm} a \equiv -{{V_{\rm A0} L}\over {2\alpha \omega}}k^2
\label{eq:xi-a}
\eeq
%
Equation~(\ref{eq:Psi}) can be rewritten in the form of Weber (parabolic cylinder)
equation (Abramowitz \& Stegun \cite{abramowitz})
\beq
{{d^2 \Psi_i}\over {d \xi^2}} + \left({\xi^2\over 4}-a
\right)\Psi_i = 0\,.
\label{eq:weber}
\eeq
Standard solutions to this equation are called Weber (parabolic
cylinder) functions, $W(a,\pm \xi)$ (Abramowitz \& Stegun \cite{abramowitz})
\beq
W(a,\pm \xi)={({\cosh {\pi a})^{1/4}}\over
{2\sqrt{\pi}}}\left(G_{\rm 1} y_{\rm 1}(\xi) \mp {\sqrt {2}G_{\rm
3}y_{\rm 2}(\xi)}\right)\, ,
\eeq
where
\beq
G_{\rm 1} = \left\vert \Gamma\left({1\over 4}+{{ia}\over 2}\right)\right\vert,\hspace{3mm}
G_{\rm 3} = \left\vert\Gamma\left ({3\over 4}+{{ia}\over 2}\right )\right\vert
\eeq
and $y_{\rm 1}(\xi),\,\,y_{\rm 2}(\xi)$ are respectively even and odd
solutions to Eq.~(\ref{eq:weber}) such as
\beqa\nonumber
y_{\rm 1}(\xi)=1+a{{\xi^2}\over {2!}}+\left (a^2-{1\over 2}\right ){{\xi^4}\over
{4!}}+\ldots \, ,
\\
\, y_{\rm 2}(\xi)=\xi+a{{\xi^3}\over {3!}}+\left (a^2-{3\over 2}\right
){{\xi^5}\over {5!}}+\ldots \, .
\eeqa

In the case of a weakly inhomogeneous Alfv\'en speed , i.e.
$\alpha^2 \ll 1$, Eq.~(\ref{eq:weber}) has periodic solutions when
\beq
a < 0,\,\,\,-a\gg \xi^2,\,\,\,p\equiv\sqrt{-a}\, .
\eeq
%
We adopt the following expansion (Abramowitz \& Stegun
\cite{abramowitz}; see also Zaqarashvili and Murawski
\cite{Zaqarashvili20072}):
\beqa
\nonumber
W(a,\xi)+i W(a,-\xi)= \\
\sqrt{2}W(a,0)\exp{[v_r + i(p \xi + {\pi/4} +v_i)]}\, ,
\label{eq:expand}
\eeqa
where:
\beqa
W(a,0)    &=& {1\over 2^{3/4}}\sqrt{{G_{\rm 1}}\over {G_{\rm 3}}}\,,\\
v_{\rm r} &=& -{{(\xi/2)^2}\over {(2p)^2}}+{{2(\xi/2)^4}\over {(2p)^4}}+\ldots \,,\\
v_{\rm i} &=& {{2/3(\xi/2)^3}\over {2p}} + \ldots \, .
\eeqa
As a result of relation $-a\gg \xi^2$ we have from Eq.~(\ref{eq:expand})
\beqa
W(a,\xi)=\sqrt{2}W(a,0)\exp{\left (-{{\xi^2}\over {16p^2}}\right)}\cos{\zeta}\, ,
\label{eq:wax}
\\
W(a,-\xi)=\sqrt{2}W(a,0)\exp{\left (-{{\xi^2}\over {16p^2}}\right
)}\sin{\zeta}\, ,\\
\zeta\equiv p \xi + {\pi/4} +{{\xi^3}\over {24p}}\, .
\eeqa
The general solution to Eq.~(\ref{eq:weber}) is
\beq \Psi_i=c_{\rm 1}W(a,\xi)+c_{\rm 2}W(a,-\xi)\,,
\label{eq:gen}
\eeq
where $c_{\rm 1}$ and $c_{\rm 2}$ are constants.

Equation~(\ref{eq:gen}) describes a periodic function and therefore
it may easily satisfy the line-tying boundary conditions of
Eq.~(\ref{eq:Psibound}). It is noteworthy that there are two sets
of solutions of Eq.~ (\ref{eq:gen}). For one set there is no node at the
loop apex i.e. at $\xi=0$. These are odd solutions which imply $c_{\rm 1}=c_{\rm 2}$ and correspond to the first, third, etc standing modes.
For the other set there is a node at $\xi=0$ and the solutions imply $c_{\rm 1} = -c_{\rm 2}$. They correspond to the second, fourth, etc standing modes. Both sets satisfy the boundary conditions of Eq.~(\ref{eq:Psibound}).

%
In the case of odd solutions, the line-tying boundary conditions of
Eq.~(\ref{eq:Psibound}) determine $k_{\rm n}$,
\beq k^2_{\rm n} - {{(2n+1)\pi}\over {2L}}k_{\rm n} + {\alpha^2
\over 6}{\omega^2 \over V^2_{\rm A0}}=0, \hspace{3mm}
n=0,1,2,\ldots\, .
\label{eq:k}
\eeq
We have
\beq
k^2_{\rm n} = {{(2n+1)^2\pi^2}\over {8L^2}}\left (1+\sqrt{1 -
{\alpha^2 \over 3}{\omega^2 \over V^2_{\rm A0}}{{8L^2}\over
{(2n+1)^2\pi^2}}} \right )-{\alpha^2 \over 6}{\omega^2
\over V^2_{\rm A0}}.
\label{eq:k2}
\eeq
Using expression (\ref{eq:k2}) we rewrite Eq.~(\ref{eq:Phi}) inside and outside the loop as
\beqa
{{d^2 \Phi_i}\over {d x^2}} - m^2_i \Phi_i=0\, ,
\label{eq:Psi1}
\\
{{d^2 \Phi_e}\over {d x^2}} - m^2_e \Phi_e=0\, ,
\label{eq:Phi1}
\eeqa

where
\beq
m^2_{\rm i} = k^2_n-{\omega^2 \over V^2_{\rm A0}}\,
\label{eq:mio}
\eeq
and
\beq
m^2_{\rm e} = k^2_n-{\omega^2 \over V^2_{\rm Ae}}.\,
\label{eq:meo}
\eeq

%
For even solutions the line-tying conditions give
\beq k^2_{\rm n} - {{n\pi}\over {L}}k_{\rm n} + {\alpha^2 \over
6}{\omega^2 \over V^2_{\rm A0}}=0, \hspace{3mm} n=1,2,3, \ldots \, ,
\eeq
which leads to
\beq k^2_{\rm n} = {{n^2\pi^2}\over {2L^2}}\left (1+\sqrt{1 -
{\alpha^2 \over 3}{\omega^2 \over V^2_{\rm A0}}{{2L^2}\over
{n^2\pi^2}}} \right ) - {\alpha^2 \over 6}{\omega^2 \over V^2_{\rm
A0}}. \eeq
%

\begin{figure}[t]
\vspace*{1mm}
\begin{center}
\includegraphics[width=7.5cm]{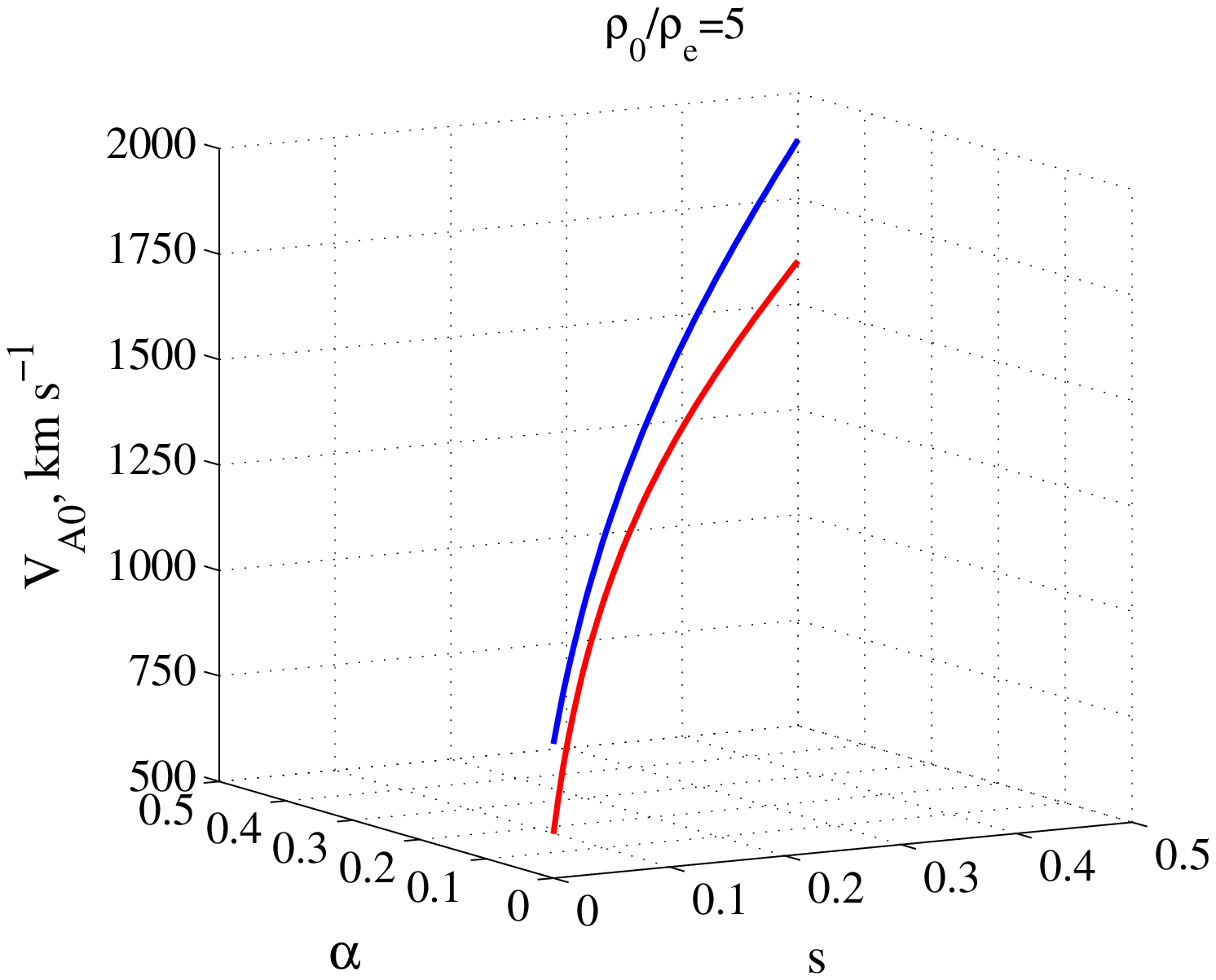}
\includegraphics[width=7.5cm]{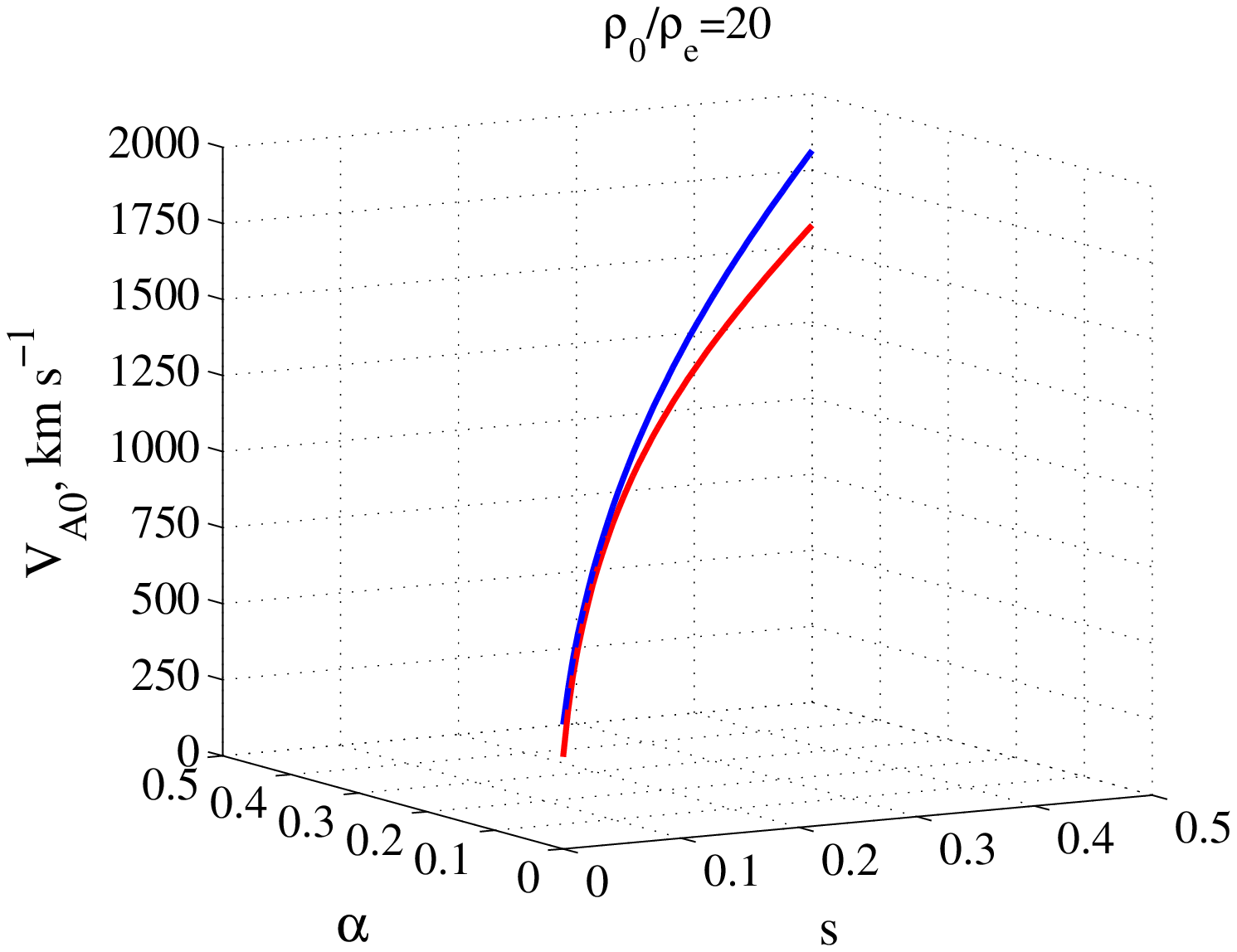}
\end{center}
\caption{Estimated Alfv\'en speed vs aspect ratio, $s=x_0/L$, and inhomogeneity parameter $\alpha$ of coronal loops. Blue (red) line corresponds to the first (second) harmonic of kink waves \textbf{according to Eq.~\ref{eq:first} (Eq.~ \ref{eq:second})}. Upper panel corresponds to ${\rho}_{\rm 0}/{\rho}_{\rm e}$=5 and the lower panel corresponds to ${\rho}_{\rm 0}/{\rho}_{\rm e}$=20. Only one line from each surfaces of $V_{A0}(s,\alpha)$ is shown here.  }
\end{figure}

\begin{figure}[t]
\vspace*{1mm}
\begin{center}
\includegraphics[width=8cm]{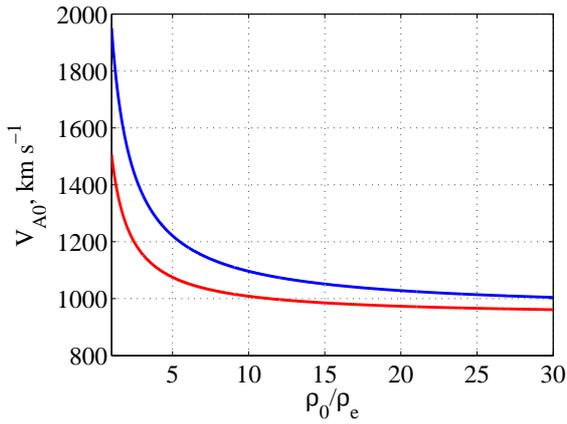}
\end{center}
\caption{Estimated Alfv\'en speed vs density contrast. Blue and red lines correspond to the Alfv\'en speeds estimated from the first and the second harmonics of kink waves, respectively.}
\end{figure}


%


Solution of transverse velocity inside the slab from Eq.~(\ref{eq:Psi1}) is
\beqa
\Phi_i=b_1 \cosh m_i x +b_2 \sinh m_i x,
\eeqa
where $b_1$ and $b_2$ are arbitrary constants.

On the other hand, outside the slab the solutions are
\beqa
\Phi_e=d_1 e^{-m_e(x-x_0)},\, {\rm for}\,\, x>x_0,
\eeqa
\beqa
\Phi_e=d_2 e^{m_e(x+x_0)},\, {\rm for}\,\, x<-x_0.
\eeqa
The solutions outside the slab must vanish at infinity, which means that $m_e>0$. Additionally, we consider a trapped solution, which means that the oscillations should be confined inside the slab. Therefore, we should avoid the oscillatory behavior outside the loop i.e. leaky modes. Hence, $m_e$ should be real value i.e. $m^2_e>0$. This condition leads to the restriction on wave frequency, which for small $\alpha$ reads as (providing that $\omega>0$)
\beqa
\omega < {{(2n+1)\pi V_{\rm Ae}}\over {2L}}\left ({1+{{\alpha^2 V^2_{Ae}}\over {3V^2_{A0}}}}\right )^{-1/2},
\label{eq:cutoffo}
\eeqa
for the odd modes and
\beqa
\omega < {{n\pi V_{\rm Ae}}\over {L}}\left ({1+{{\alpha^2 V^2_{Ae}}\over {3V^2_{A0}}}}\right )^{-1/2}
\label{eq:cutoffe}
\eeqa
for even modes.

In order to obtain the dispersion relation, we need the expression of total pressure perturbations which is readily obtained
\beqa
P={{B_{\rm 0}b_z}\over 4\pi}=-i{{B^2_{\rm 0}}\over {4\pi \omega}}{{\partial u_x}\over {\partial x}}.
\eeqa

Continuity of transverse velocity and total pressure perturbations at slab boundaries $x=\pm x_0$ gives the dispersion relations
for even (with respect to $x$)
\beqa
m_i \sinh m_i x_0+m_e \cosh m_i x_0=0,
\label{eq:even}
\eeqa
and odd (with respect to $x$) modes
\beqa
m_i \cosh m_i x_0+m_e \sinh m_i x_0=0.
\label{eq:odd}
\eeqa
Odd mode has velocity node at the tube center, $x=0$, and it is equivalent to sausage waves, while even mode has no velocity node at $x=0$ and it is equivalent to kink waves. Kink waves lead to the transverse displacement of slab axis, while sausage modes do not (Roberts \cite{Roberts1981}).

\subsection{Estimation of plasma parameters in the outer solar corona}

Eq.~(\ref{eq:even}) can be solved analytically for $m_ix_0 \ll 1$, which yields $\sinh m_i x_0 \approx m_i x_0$ and $\cosh m_i x_0 \approx 1$. Then the frequency of the first harmonic of kink waves can be approximated as (assuming $s$, $\alpha$ and $\rho_{\rm e}/{\rho}_{\rm 0}$ to be small)
\beqa
\omega_{1k} \approx \sqrt{{{2 +\pi s}\over {\pi(1+{\alpha^2/3})s+{\rho}_{\rm e}/{\rho}_{\rm 0}}}}{{\pi V_{\rm A0}}\over {2L}},
\label{eq:first}
\eeqa
where $s=x_0/L$ is the ratio of loop width and length, while the frequency of the second harmonic is
\beqa
\omega_{2k} \approx \sqrt{{{2(1 +\pi s)}\over {2\pi(1+{\alpha^2/3})s+{\rho}_{\rm e}/{\rho}_{\rm 0}}}}{{\pi V_{\rm A0}}\over {L}}.
\label{eq:second}
\eeqa

If the observed periods of radio emission intensity variation with $34$ min and $23$ min correspond to the first and second harmonics of kink waves, then  Eqs.~(\ref{eq:first})-(\ref{eq:second}) allow to estimate the value of Alfv\'en speed $V_{\rm A0}$ as well as the values of $s$ and $\alpha$. The Alfv\'en speed should be the same as estimated from both harmonics, therefore we should find the values of $s$ and $\alpha$, in which the condition is satisfied. Fig. 8 shows that the curves corresponding to the estimated Alfv\'en speed for the first and second harmonics do not intersect in the case of $\rho_{\rm 0}/\rho_{\rm e}=5$ (here we use the loop length of $2L=2.64 \cdot 10^6$ km). Therefore, the observations can not be explained by kink oscillations in this case. On the other hand, the curves approach each other for $\rho_{\rm 0}/\rho_{\rm e}=20$ if
\beqa
s={x_0\over L}\approx 0.05-0.1
\label{eq:salpha}
\eeqa
and
\beqa
\alpha \approx 0.05-0.1.
\label{eq:salpha}
\eeqa
In this case, the Alfv\'en speed in the intersection region ($\alpha$=0.1, $s$=0.1) is of the order of
\beqa
V_{\rm A0}\approx 1000 \, {\rm km}{\cdot}{\rm s}^{-1}.
\label{eq:alfe}
\eeqa

Figure 9 shows the estimated Alfv\'en speed vs the density contrast for $\alpha$=0.1 and $s$=0.1 according to Eqs.~(\ref{eq:first})-(\ref{eq:second}). It is clearly seen that the Alfv\'en speeds estimated from the first and the second harmonics are very different for small density contrast, but become relatively similar for higher density contrast and tend to $\sim$ 1000 km s$^{-1}$.

We solved the dispersion relation for kink waves (Eq.~(\ref{eq:even})) numerically. The ratio of the frequencies of first and second harmonics vs the ratio of internal to external densities is plotted on Figure 10. The ratio increases for large $\rho_{\rm 0}/\rho_{\rm e}$. On the other hand, the ratio of observed oscillation periods is $23/35 \sim 0.65$. Therefore, it is seen that $\rho_{\rm 0}/\rho_{\rm e}$ should be et least $\geq 20$ in order to explain the observations. This means that the transequatorial loops are much denser than the surrounding plasma in the outer corona. However, this estimation was derived on the base of calculations for a straight coronal slab. Consideration of more realistic situation of curved loop with the tube geometry may significantly change the estimation.

The value of Alfv\'en speed (Eq.~\ref{eq:alfe}) is estimated near the height of 1 $R_0$ which is the averaged altitude of transequatorial loop apex. Then the Alfv\'en speed at the coronal base can be estimated from Eq.~(\ref{eq:rhoi}) for $\alpha=0.1$ as
\beqa
V_{\rm A0}\approx 995 \, {\rm km}{\cdot}{\rm s}^{-1}.
\label{eq:alfe2}
\eeqa
Therefore, the Alfv\'en speed stays almost unchanged from coronal base up to the height of $\sim R_0$.

\begin{figure}[t]
\vspace*{1mm}
\begin{center}
\includegraphics[width=8.5cm]{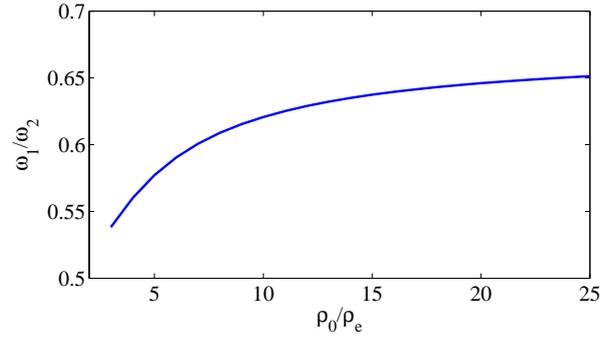}
\end{center}
\caption{Ratio of frequencies of first and second harmonics vs the ratio of internal to external densities. The aspect ratio is taken as $x_0/L$=0.1 and the inhomogeneity parameter as $\alpha=0.1$.}
\end{figure}

Alfv\'en speed and electron number density allow us to estimate the magnetic field strength inside the loop in the outer corona. Electron number density can be obtained from observed radio frequencies according to Eq.~\ref{plasma-frequency} (the radio frequencies of 30-14 MHz correspond to the averaged value of electron number density of 4 $\cdot$10$^{6}$ cm$^{-3}$). Note that the same value is estimated from gravitationally stratified corona at the height of 1 $R_0$. Then, Eq.~(\ref{eq:alfe}) gives the value of mean magnetic field strength as
\beqa
B_0 \approx 0.9 \, {\rm G}.
\label{eq:B}
\eeqa


\section{Discussion}

Observed oscillations are important tools to estimate the plasma parameters in the solar atmosphere. Up to now, the oscillations have been used for the inner corona as a coronal seismology (Nakariakov and Ofman \cite{Nakariakov2001}), for prominences as prominence seismology (Arregui et al. \cite{Arrequi2012}) and for solar spicules as a spicule seismology (Zaqarashvili and Erd\'elyi \cite{Zaqarashvili2009}). In the case of coronal loops, the observed oscillations in EUV line intensity may allow to estimate the magnetic field strength in the corona up to 0.1 $R_0$. In other cases, the oscillations in H$\alpha$ line may allow to estimate the magnetic field strength in chromospheric spicules and prominences. The coronal loops are not seen in the outer corona ($>$ 0.2 $R_0$) because of the decrease in EUV intensity. Therefore, the coronal seismology does not allow to estimate the plasma parameters in the outer corona. On the other hand, radio emissions at the frequency range of 10-100 MHz, which corresponds to the heights of 0.2-1.3 $R_0$, can be successfully used for the radio sounding of these heights.

We analyzed the type IV radio burst at the frequency range of 13-32 MHz which corresponds to the heights of 0.6-1.3 $R_0$ as estimated from Baumbach-Allen formula and to the heights of 0.7-1.2 $R_0$ as estimated from the gravitationally stratified corona for 1 MK temperature (Fig. 2). The type IV radio burst was connected to small C2.3 flare. The burst is believed to be caused by plasma mechanism of radio emission from Langmuir waves, which are triggered by energetic electron beams after the solar flare. The electrons are accelerated along the coronal loops, therefore the radio emission probably originates from the loops. There are two possible kinds of loops, which may expand up to the observed heights. The active region loops, which connect the opposite polarities in the same active region, may expand up to higher corona. Consequently the closed loops may reach the base of helmet streamers, where the magnetic field lines start to open. On the other hand, the transequatorial loops, which connect the active regions of opposite hemispheres, may reach the observed heights, because of large distance between their footpoints. At least two different transequatorial loop systems may exist on the visible part of the Sun during the observed type IV burst (see Fig. 4). One of them is connected to the active region, where the flare occurred. Therefore we suggest that the observed radio emission was originated from the transequatorial loop system, which connects the active regions AR11190 and AR11187 (see Fig. 6).

We applied the wavelet analysis to study the periodic variations of type IV burst intensity at four frequencies: 29 MHz, 25 MHz, 22 MHz and 14 MHz. The analysis shows two clear periodicities in the emission intensity with significant probability (Fig. 3). The period of $\sim$ 34 min has stronger peak in the power spectrum and it appears at all frequencies of radio emission. The period of $\sim$ 23 min appears only at frequencies of 25 MHz, 22 MHz and it is absent at 29 MHz and 14 MHz frequencies. We suggest that the periodicity in the emission intensity can be caused by MHD oscillations of coronal loops, which were triggered by the same flare. The apex of the coronal loop is located at the height of 1.2 $R_0$ using the estimations from both, Baumbach-Allen formula (Eq.~\ref{baumbach-allen}) and the gravitationally stratified atmosphere with the temperature of 1 MK (Eq.~\ref{strat}). The possible length of the loop and consequently  the phase speed of 34 min oscillations can be estimated as 2.64 $\cdot 10^6$ km and 2500 km $\cdot$ s$^{-1}$, respectively, if the fundamental harmonic is concerned. Therefore, we may rule out the slow magneto-acoustic mode as the reason of the periodicity because of its low phase speed. There are still three MHD modes, which could be responsible for the oscillations: kink, sausage and torsional Alfv\'en modes. The sausage waves have long wavelength cut-off in magnetic tubes, so that the first and second harmonics seem to be evanescent in typical coronal loops. The third harmonic can be trapped in some conditions, however the estimation gives rather small value of Alfv\'en speed in the corona (see subsection 4.2). Therefore, we think that the sausage waves can be also ruled out. It is hard to excite torsional oscillations in coronal loops, therefore fast kink waves are best candidates for explaining the observed periodicity. There are two types of kink modes in coronal loops: the horizontal kink mode, which displaces the loop horizontally and the vertical kink mode, which displaces the loop vertically. Horizontal kink oscillations are frequently observed in the solar corona. They are almost incompressible, hence do not lead to significant variation in plasma density. The oscillations damp quickly over few wave periods. On the other hand, vertical kink oscillations are more rarely observed. Recently, Aschwanden and Schrijver (\cite{aschwanden2011}) found \textbf{a} very interesting example of vertical kink oscillations. The oscillations were accompanied with variation of cross section/density and they showed no damping during the interval of four oscillation periods. They suggested that the variation of density and cross section is the result of the variation of loop length during the oscillation. In our case, the oscillation is accompanied by variation of cross section (because it should modulate the electron beam density) and it does not show the damping during several wave periods. Therefore, we think that the oscillation, which resulted in the variation of radio emission intensity in type IV burst, has the properties similar to that reported by Aschwanden and Schrijver (\cite{aschwanden2011}), hence it is the vertical kink oscillation of transequatorial loop.

%

Observation of two different periods in the emission intensity (34 min, 23 min) can be explained by the first and second harmonics of standing oscillations. Their ratio is significantly shifted from 0.5 being $\sim$ 0.67, which may cause some problems at the first glance. However, the oscillations in the magnetic tubes indeed have the property that the ratio is different than 0.5. There are several observational examples of simultaneous existence of first and second harmonics of kink (Van Doorsselaere et al. \cite{van1}, Verwichte et al. \cite{verwichte}, De Moortel \& Brady \cite{demoo}) and sausage (Srivastava et al. \cite{Srivastava}) oscillations in coronal loops. The ratio between the frequencies of first and second harmonics was significantly shifted from 0.5 in all observations, which was explained as a result of longitudinal density stratification in the loop (Andries et al. \cite{andries2}, McEwan et al. \cite{mc}). It should be mentioned, however, that the finite radius of the tube and the density contrast inside and outside the tube also lead to the shift of the ratio from 0.5. We solved MHD equations for a coronal slab with longitudinally inhomogeneous Alfv\'en speed and derived the dispersion relation for kink and sausage waves. From the observed frequency ratio we then found that the value of the inhomogeneity parameter $\alpha$ and the aspect ratio (the ratio of loop width and length) should be around 0.1, while the density contrast should be large enough, $\rho_{\rm 0}/\rho_{\rm e}>20$. Therefore, the analysis is in favor for a dense loop with weak longitudinal inhomogeneity of the Alfv\'en speed. However, the analysis was performed for a straight coronal loop with the slab geometry. On the other hand, more real situation of curved coronal loop with the tube geometry may significantly change the situation.

The Alfv\'en speed inside the loop at its apex (approximately 1 $R_0$) for $\alpha=0.1$, $x_0/L=0.1$ and $\rho_{\rm 0}/\rho_{\rm e}=20$ is estimated as $\sim$ 1000 km $\cdot$ s$^{-1}$. Then the magnetic field strength inside the loop can be derived fixing the value of the loop density. Electron number density inside the loop can be obtained from observed radio frequencies of 14-29 MHz which correspond to the averaged value of electron number density of 4 $\cdot$10$^{6}$ cm$^{-3}$. Note that the same value is estimated from the gravitationally stratified corona with 1 MK temperature at the height of 1 $R_0$. Using the value of loop density we may estimate the mean magnetic field strength at the height of 1 $R_0$ as $\sim$ 0.9 G. We then may extrapolate the Alfv\'en speed towards the loop footpoints using $\alpha=0.1$ and it gives $\sim$ 995 km $\cdot$ s$^{-1}$. Therefore, the Alfv\'en speed does not significantly vary with height up to 1 solar radius, hence it yields $B(z)\sim \sqrt{\rho(z)}$. However, estimated density contrast $\rho_{\rm 0}/\rho_{\rm e}=20$ seems to be high at these heights. For the electron number density of 4 $\cdot$10$^{6}$ cm$^{-3}$ inside the loop (estimated from radio emission frequencies), the density contrast leads to the electron number density outside the loop as 2 $\cdot$10$^{5}$ cm$^{-3}$. On the other hand, the Baumbach-Allen formula estimates the electron number density outside the active region as $\sim$ 3.7 $\cdot$10$^{6}$ cm$^{-3}$ at 1 $R_0=1$, which is much higher than the estimated value. Requested value of external density can be obtained from Baumbach-Allen formula only near 5 $R_0$, which seems to be much higher height for coronal loops. Therefore, either the Baumbach-Allen formula is not valid at the heights of 1 $R_0$ or the estimations require further revision. The Baumbach-Allen formula has been tested in different situations and it seems to work quite well in the solar corona. Therefore, we think that more realistic model of curved coronal tube may significantly reduce the density contrast and remove the inconsistency. Corresponding calculations should be done in future.

The constancy of Alfv\'en speed with height and the electron number densities at the heights of 0.1 $R_0$ and 1 $R_0$ (4.67 $\cdot$ 10$^8$ cm$^{-3}$ and 4$\cdot$ 10$^6$ cm$^{-3}$ from gravitationally stratified corona) allows to estimate the magnetic field strength at 0.1 $R_0$ as $\sim$ 10 G. Note, that early estimations of magnetic field strength in coronal loops at the height of 0.1-0.2 $R_0$ are 300 G by Asai et al. (\cite{Asai2001}), 13$\pm$9 by Nakariakov and Ofman (\cite{Nakariakov2001}), 2-51 G by Wang et al. (\cite{Wang2007}) and $\sim$ 4 G by Aschwanden and Schrijver (\cite{aschwanden2011}). The magnetic field strength estimated from our observations agrees with the estimation of Nakariakov and Ofman (\cite{Nakariakov2001}).

The density structure of gravitationally stratified atmosphere with 1 MK temperature has \textbf{a} scale height of $\sim$ 0.086 $R_0$ near the coronal base. The constancy of Alfv\'en speed with height gives the scale height of magnetic field strength to be $\sim$ 0.17 $R_0$. On the other hand, if one considers a current-free magnetic arcade, then, the $z$-dependence of magnetic field strength is proportional to $\exp(-z/\Lambda_B)$ with $\Lambda_B=l/\pi$, where $l$ is the horizontal extent of the arcade. For transequatorial loops one may consider $l$ as the distance between active regions in opposite hemispheres i.e. in our case between AR11190 and AR11187, which is about 0.7 $R_0$. This gives $\Lambda_B\approx$ 0.22 $R_0$. Therefore, seismologically estimated scale height is slightly different than that of current-free magnetic configuration. This may indicate that the magnetic structure is not current-free and the current is caused by shearing and/or twisting of arcade due to footpoint motions.

This is the first attempt to use the radio observations for estimation of plasma parameters in the outer solar corona ($>0.2 R_0$). More observations and theoretical modeling are needed to test the estimations and to develop the seismology of outer solar corona.

\section{Conclusion}

We start using the radio observations to estimate the plasma parameters and to develop the {\it radio seismology} of the outer solar corona ($>0.2 R_0$). As an example, we use the observations from Ukrainian radio telescope URAN-2 in April 14, 2011. Type IV radio burst was detected in the frequency range of 13-32 MHz during the time interval of 09:50-12:30 UT, which is probably connected to small C2.3 flare occurring in the active region AR 11190 between 09:38-09:49 UT. Wavelet analysis at four different frequencies (29 MHz, 25 MHz, 22 MHz and 14 MHz) shows the periodicity of 34 min and 23 min in the emission intensity. We suggest that the periodicity is caused by the first and second harmonics of vertical kink oscillations of transequatorial loop system, which connects the active regions from northern and southern hemispheres. Using the theoretical properties of kink oscillations in longitudinally inhomogeneous coronal loops, we estimate the Alfv\'en speed inside the loop at the height of 1 $R_0$ as $\sim$ 1000 km s$^{-1}$. The coronal loop seems to be thin (with width to length ratio of 0.1), longitudinally weakly inhomogeneous and dense ($\rho_{\rm 0}/\rho_{\rm e} \geq 20$). Estimated magnetic field strength of the loop at the height of 1 $R_0$ is $\sim$ 0.9 G. Extrapolation of this value gives the magnetic field strength at the height of 0.1 $R_0$ as $\sim$ 10 G, which is in agreement with the estimation of Nakariakov and Ofman (\cite{Nakariakov2001}). More realistic calculations of MHD oscillations in a curved coronal tube with a longitudinal inhomogeneity may increase the accuracy of the estimations.

\begin{acknowledgements}
The work was supported by the European FP7-PEOPLE-2010-IRSES-269299 project- SOLSPANET. The work of TZ was also supported by EU collaborative project STORM - 313038 and by Shota Rustaveli National Science Foundation grant DI/14/6-310/12. The work of MP was supported by the Austrian Fonds zur F\"orderung der wissenschaftlichen Forschung (project P23762-N16).
\end{acknowledgements}

\appendix
\section{Density stratification in the solar atmosphere}

Equilibrium pressure and density in a stratified atmosphere under gravity satisfy the equation
\beqa
{{dp}\over {dR}}=-\rho(R){{GM}\over {R^2}},
\label{eq:strat}
\eeqa
where $G=6.674 \cdot 10^{-8}$ cm$^3$ g$^{-1}$ s$^{-2}$ is the gravitational constant, $M=1.989\cdot 10^{33}$ g is the solar mass and $R$ is the distance from the solar center.

We may integrate Eq.~\ref{eq:strat} to obtain
\beqa
p(R)=p_0(R_0)e^{-S_0(R)}, \,\, \rho(R)=\rho_0(R_0)e^{-S_0(R)},
\label{eq:p}
\eeqa
where $R_0$ is the solar radius,
\beqa
S_0(R)= \int_{R_0}^{R}{{dz}\over {H_0(z)}}
\label{eq:S}
\eeqa
is the integrated scale height and
\beqa
H_0(R)= {{p_0(R_0)R^2}\over {GM \rho_0(R_0)}}
\label{eq:H}
\eeqa
is the pressure scale height. If we consider the equation of state $p=2k\rho T/m$, where $k=1.38\cdot 10^{-16}$ erg K$^{-1}$ is the Boltzmann constant, $m=1.67 \cdot 10^{-24}$ g is the proton mass and $T=const$ is the temperature, then we obtain from Eq.~\ref{eq:S}
\beqa
S_0(R)= {{GM m}\over {2 k T R_0}}{{R-R_0}\over {R}}.
\label{eq:S1}
\eeqa
Now considering $R=R_0+h$ we derive the following formula for the density
\beqa
\rho(h)=\rho_0e^{-{R_0\over H_n}{h\over {R_0+h}}},
\label{eq:S1}
\eeqa
where $H_n=2kTR^2_0/G M m$ is the pressure scale height near the solar surface. Then we obtain the equation for electron number density as
\beqa
n_e(\zeta)=n_{e0}e^{-{R_0\over H_n}{\zeta\over {1+\zeta}}},
\label{eq:S1}
\eeqa
where $\zeta=h/R_0$ is the normalized height above the solar surface.


\begin{thebibliography}{}

\bibitem[1964]{abramowitz}Abramowitz, M., \& Stegun, I.A. 1964,
{\it Handbook of Mathematical Functions} (Washington, D.C.: National
Bureau of Standards)

\bibitem[2005a]{andries1}Andries, J., Goossens, M., Hollweg, J. V., Arregui, I., \&  Van Doorsselaere, T. 2005a,
A\&A, 430, 1109

\bibitem[2005b]{andries2}Andries, J., Arregui, I. \& Goossens, M. 2005b, \apj, 624, L57

\bibitem[2009]{Andries2009}Andries, J., van Doorsselaere, T., Roberts, B., Verth, G., Verwichte, E. and Erd\'elyi, R., 2009, Space Science Reviews, 149, 3

\bibitem[2007]{Arrequi2007}Arregui, I., Andries, J., Van Doorsselaere, T., Goossens, M. and Poedts, S., 2007, A\&A, 463, 333

\bibitem[2011]{Arrequi2011}Arregui, I. and Asensio Ramos, A., 2011, \apj, 740, 44, 10

\bibitem[2012]{Arrequi2012}Arregui, I., Oliver, R. and Ballester, J.L., 2012, Living Reviews in Solar Physics, 9, http://www.livingreviews.org/lrsp-2012-2

\bibitem[2001]{Asai2001}Asai, A., Shimojo, M., Isobe, H., Morimoto, T., Yokoyama, T.,
Shibasaki, K., \& Nakajima, H. 2001, ApJ, 562, L103

\bibitem[1999]{aschwanden1}Aschwanden, M.J., Fletcher, L., Schrijver, C.J., \& Alexander, D. 1999, \apj, 520, 880

\bibitem[2004]{Aschwanden2004}Aschwanden, M.J., 2004, Physics of the Solar Corona (Praxis/Springer, Chichester/New York)

\bibitem[2004]{Aschwanden20041}Aschwanden M. J., Nakariakov V. M., Melnikov V. F., 2004, ApJ, 600, 458

\bibitem[2011]{aschwanden2011}Aschwanden, M.J. and Schrijver, C.J., 2011, \apj, 736, 102, 20

\bibitem[2005]{brady}Brady C.S., \& Arber T.D. 2005, A\&A, 438, 733

\bibitem[2005]{Brazhenko2005}Brazhenko, A. I., Bulatsen, V. G., Vashchishin, R. V., Frantsuzenko, A. V., Konovalenko, A. A., Falkovich, I. S., Abranin, E. P., Ulyanov, O. M., Zakharenko, V. V., Lecacheux, A. and Rucker, H.O., 2005, Kinematika i Fizika Nebesnykh Tel, 5, 43

\bibitem[1976]{Chase1976}Chase, R.C., Krieger, A.S., Svestka, Z., \& Vaiana, G.S., 1976, in Space
Research XVI, (Berlin : Akademie), 917

\bibitem[2007]{demoo}De Moortel I., \& Brady, C.S. 2007, A\&A, 664,
1210

\bibitem[2004]{diaz1}D\'iaz, A.J., Oliver, R., Ballester, J.L. \& Roberts, B. 2004, A\&A, 424,
1055

\bibitem[2006]{diaz2006}D\'iaz A. J., Zaqarashvili T. V., Roberts B., 2006, A\&A, 455, 709

\bibitem[2006]{donnelly}Donnelly, G.R., Di\'az, A., \& Roberts, B. 2006, A\&A, 457, 707

\bibitem[2010]{Dorovsky2010}Dorovsky, V. V., Melnik, V. N., Konovalenko, A.A., Rucker, H.O., Abranin, E. P. and Lecacheux, A., 2010, Radio Physics and Radio Astronomy, 1, 181

\bibitem[2006]{dymova}Dymova, M. V., \& Ruderman, M. S. 2006, A\&A, 457, 1059

\bibitem[1983]{Edwin1983}Edwin P.M. \& Roberts B., 1983, Sol. Phys., 88, 179

\bibitem[1958]{Ginzburg1958}Ginzburg, V. L. and Zhelezniakov, V. V., 1958, Astronomicheskii Zhurnal, 35, 694

\bibitem[2008]{Gruszecki}Gruszecki, M., Murawski, K. and Ofman, L., 2008, A\&A, 488, 757

\bibitem[2011]{Khodachenko2011}Khodachenko, M. L., Kislyakova, K. G., Zaqarashvili, T. V., Kislyakov, A. G., Panchenko, M., Zaitsev, V. V., Arkhypov, O. V. and Rucker, H. O., 2011, A\&A, 525, A105, 4
	
\bibitem[2012]{Lemen2012}Lemen, J.R., Title, A.M., Akin, D.J. et al. 2012, Solar Phys., 275, 17

\bibitem[2006]{mc}McEwan, M. P., Donnelly, G. R., Diaz, A. J. \& Roberts, B. 2006, A\&A, 460, 893

\bibitem[2008]{McEwan2008}McEwan, M. P., D\'iaz, A. J. and Roberts, B., 2008, A\&A, 481, 819

\bibitem[2003]{Megn2003}Megn, A.V., Sharykin, N.K., Zaharenko, V.V., Bulatsen, V.G., Brazhenko, A.I. and Vachishin, R. V., 2003, Radio Physics and Radio Astronomy, 8, 345

\bibitem[2010]{Melnik2010}Melnik, V. N., Rucker, H. O., Konovalenko, A. A., Dorovskyy, V. V., Abranin, E. P. and Lecacheux, A., 2011, in Planetary Radio Emissions VII, Austrian Academy of Sciences Press, Vienna, 343

\bibitem[2008]{Melnik2008}Melnik, V. N., Rucker, H. O., Konovalenko, A. A., Dorovskyy, V. V., Abranin, E. P., Brazhenko, A.I., Thide, B., Stanislavskyy, A. A., 2008,  ed. Pingzhi Wang, Nova Science Publishers, New York, 287]

\bibitem[1999]{nakariakov1}Nakariakov, V.M., Ofman, L., Deluca, E.E., Roberts, B., \& Davila, J.M. 1999, Science, 285, 862

\bibitem[2001]{Nakariakov2001}Nakariakov V. M. \& Ofman L., 2001, A\&A, 372, L53

\bibitem[2003]{Nakariakov2003}Nakariakov V. M., Melnikov V. F. \& Reznikova V. M., 2003, A\&A, 412, L7

\bibitem[2005]{nakariakov4}Nakariakov, V.M., \& Verwichte, E. 2005, Living Reviews in Solar Physics, 2, http://www.livingreviews.org/lrsp-2005-3

\bibitem[1999]{ofm1}Ofman, L., Nakariakov, V. M., \& DeForest, C. E. 1999, ApJ, 514, 441

\bibitem[2009]{Pascoe2009}Pascoe, D. J., Nakariakov, V. M., Arber, T. D. and Murawski, K., 2009, A\&A, 494, 1119


\bibitem[2011]{Pesnell2011}Pesnell, W. D., Thompson, B.J. and Chamberlin, P. C., 2011, Solar Phys., 275, 3

\bibitem[2000]{Pevtsov2000}Pevtsov, A.A., 2000, \apj, 531, 553

\bibitem[1981]{Roberts1981}Roberts, B., 1981, Solar Phys., 69, 39

\bibitem[1984]{roberts1984}Roberts B., Edwin P.M. \& Benz A.O., 1984, \apj, 279, 857

\bibitem[2010]{Ryabov2010}Ryabov, V.B., Vavriv, D.M., Zarka, P., Ryabov, B.P., Kozhin, R., Vinogradov, V.V., Denis, L., 2010, A\&A, 510, A16


\bibitem[2012]{Schou2012}Schou, J., Scherrer, P. H., Bush, R. I., et al., 2012, Solar Phys., 275, 229

\bibitem[2005]{selwa}Selwa, M., Murawski, K., Solanki, S. K., Wang, T. J., \& T\'oth, G.
2005, A\&A, 440, 385

\bibitem[2007]{Selwa2007}Selwa, M., Murawski, K., Solanki, S. K. and Wang, T. J., 2007, A\&A, 462, 1127

\bibitem[2008]{Srivastava}Srivastava, A. K., Zaqarashvili, T. V., Uddin, W., Dwivedi, B. N. and Kumar, Pankaj, 2008, MNRAS, 388, 1899

\bibitem[2012]{Sych2012}Sych, R., Zaqarashvili, T. V., Nakariakov, V. M., Anfinogentov, S. A., Shibasaki, K. and Yan, Y., A\&A, 539, A23, 10

\bibitem[2006]{terradas}Terradas, J., Oliver, R., \& Ballester, J.L. 2006, \apj, 642, 533

\bibitem[2004]{van1}Van Doorsselaere T., Debosscher A., Andries J. \& Poedts S. 2004, A\&A, 424, 1065

\bibitem[2007]{Van Doorsselaere2007}Van Doorsselaere, T., Nakariakov, V. M. and Verwichte, E., 2007, A\&A, 473, 959

\bibitem[2008]{verth}Verth G., \& Erd\'elyi R., 2008, A\&A, 486, 1015

\bibitem[2010]{Verth2010}Verth, G., Erd\'elyi, R. and Goossens, M., 2010, ApJ, 714, 1637

\bibitem[2011]{Verth2011}Verth, G., Goossens, M., and He, J.-S., 2011, ApJL, 733, L15, 5

\bibitem[2004]{verwichte} Verwichte, E., Nakariakov, V.M., Ofman, L., \& DeLuca E.E. 2004, Solar Physics, 223, 77

\bibitem[2006]{verwichte1}Verwichte, E., Foullon, C., \& Nakariakov, V. M. 2006, A\&A, 446, 1139
%
\bibitem[2003a]{wang1}Wang, T.J., Solanki, S.K., Innes, D.E. et al. 2003a, A\&A, 402, L17

\bibitem[2003b]{wang3}Wang, T.J., Solanki, S.K., Curdt, W. et al. 2003b, A\&A, 406, 1105

\bibitem[2004]{wang2}Wang, T.J., \& Solanki S.K. 2004, A\&A, 421, L33

\bibitem[2007]{Wang2007}Wang, T.J., Innes, D.E. and Qiu, J., 2007, ApJ, 656, 598

\bibitem[2003]{Zaqarashvili2003}Zaqarashvili, T.V., 2003, A\&A, 399, L15
%
\bibitem[2007]{Zaqarashvili20071}Zaqarashvili, T.V., Khutsishvili, E., Kukhianidze, V. and Ramishvili, G., 2007, A\&A, 474, 627
%
\bibitem[2007]{Zaqarashvili20072}Zaqarashvili, T.V., \& Murawski, K. 2007, A\&A, 470, 353
%
\bibitem[2009]{Zaqarashvili2009}Zaqarashvili, T.V. \& Erd{\'e}lyi, R., 2009, Space Sci. Rev., 149, 355

\end{thebibliography}
\end{document}